%% file: main_version_FINAL.tex
\documentclass[twocolumn]{aastex631}
\usepackage{CLASSY_Preamble}

\begin{document}
\submitjournal{AASJournal ApJ}
\shortauthors{Arellano-C\'ordova et al.}
\shorttitle{Gas-phase abundance patterns}
\color{magenta}
\title{CLASSY IX: The Chemical Evolution of the Ne, S, Cl, and Ar Elements}
\color{black}

\include{authors}

\correspondingauthor{Karla Z. Arellano-C\'ordova} 
\email{ziboney@gmail.com, k.arellano@ed.ac.uk}


\begin{abstract}
To study the chemical evolution across cosmic epochs, we investigate Ne, S, Cl, and Ar abundance patterns in the COS Legacy Archive Spectroscopic SurveY (CLASSY).
CLASSY comprises local star-forming galaxies (0.02 $< z <$ 0.18) with enhanced star-formation rates, making them strong analogues to high-$z$ star-forming galaxies. With direct measurements of electron temperature, we derive accurate ionic abundances for all elements and assess ionization correction factors (ICFs) to account for unseen ions and derive total abundances. We find Ne/O, S/O, Cl/O, and Ar/O exhibit constant trends with gas-phase metallicity for 12+log(O/H) < 8.5 but significant correlation for Ne/O and Ar/O with metallicity for 12+log(O/H) > 8.5, likely due to ICFs. Thus, applicability of the ICFs to integrated spectra of galaxies could bias results, underestimating true abundance ratios. Using CLASSY as a local reference, we assess the evolution of Ne/O, S/O, and Ar/O in galaxies at $z>3$, finding no cosmic evolution of Ne/O, while the lack of direct abundance determinations for S/O and Ar/O can bias the interpretation of the evolution of these elements. We determine the fundamental metallicity relationship (FMR) for CLASSY and compare to the high-redshift FMR, finding no evolution. Finally, we perform the first mass-neon relationship analysis across cosmic epochs, finding a slight evolution to high Ne at later epochs. The robust abundance patterns of CLASSY galaxies and their broad range of physical properties provide essential benchmarks for interpreting the chemical enrichment of the early galaxies observed with the JWST.     

\end{abstract} 


\section{Introduction}\label{sec:intro}

Recent observations of \textit{JWST} have opened a new window into the exploration of the chemical enrichment of galaxies at $z > 6$ \citep[e.g.,][]{schaerer22, arellanocordova22b, curti23a, brinchmann23, rhoads23, isobe23a, nakajima23}. The determination of the oxygen abundances (O/H, often referred to as metallicity) is an essential goal of \textit{JWST} observations due to its strong relationship with galaxy properties such as stellar mass and star-formation rate. Such scaling relationships involve diverse mechanisms associated with galaxy formation and evolution such as star formation, outflows, or inflows of pristine gas 
 \citep[i.e., the baryon cycle;][]{ tremonti04, tumlinso17}. Impressively, direct O/H abundances have already been computed for a small number of sources thanks to the successful detection of the temperature-sensitive emission line of [\ion{O}{3}]~\W4363 at $z>7$ \citep[e.g.,][]{schaerer22, curti23a, rhoads23, trump23, arellanocordova22b, laseter23}. While direct abundance determinations are often viewed as the most robust method, recent temperature inhomogeneities results in \hii regions may result in a strong bias in the numer of objects with high degree of ionization objects, like those observed at z $ <7$ \citep{mendez-delgado+23a, cameron23, arellanocordova22b}. Such bias might have a significant impact on important scaling relations such as the mass-metallicity relation in metal-poor objects. Samples of galaxies at $z\sim0$ with robust determinations of electron temperature (i.e. \Te, using different line diagnostics) and metallicities are crucial to assess this bias and the potential impacts on high redshift abundance determinations.

In addition to oxygen, strong emission lines associated with other elements are now being clearly detected at $z >4$ \citep[e.g.,][]{arellanocordova22b, nakajima23, isobe23a, marqueschaves23, jones23}. Example includes Ne, S, and Ar, which are the most typical $\alpha$-elements observed in nebular studies in the local Universe \citep{izotov06, croxall16}. We do not include nitrogen here because it is the focus of an upcoming paper (Arellano-C\'ordova et al. in prep.) \citep[see also,][]{stephenson23}. These elements are ejected into the interstellar medium (ISM) by core-collapse supernovae (CCSNe)  \citep[][]{henry1999, prantzos20, kobayashi20}. 

Before \textit{JWST}, the majority of $\alpha$-element abundance measurements were limited to \hii\ regions and  star-forming galaxies (SFGs) from the local universe. \citep[e.g.,][]{croxall16, arellano-cordova2020b, berg18, berg19, kumari+19a, Dominguez-Guzman2022, guseva11, gusev12, izotov06, izotov11, Rogers22, diaz22}. In general, many studies report little-to-no variation between $\alpha$-elements with respect to O and metallicity \citep[e.g.,][]{kennicutt03b, izotov06, berg21, Rogers22}, with the exception of S/O which has been seen to decrease significantly as metallicity increases \citep[e.g., S/O \textit{vs} O/H,][]{amayo21, diaz22, izotov06, guseva11}.

Recent observations with \textit{JWST} now present the opportunity to determine the chemical abundances of galaxies at high-$z$. Chemical abundances of SFGs have been studied in surveys like CEERS, GLASS, and JADES \citep{finkelstein22a, treu22,eisenstein23}. For example, the abundance ratio of C/O have been reported in a few galaxies at $z > 6$ \citep[e.g., ][]{arellanocordova22b, cameron23b, jones12,marqueschaves23}, while the chemical abundances of Ne, S, and Ar have been determined for multiple SFGs at $z =4-8$ \citep[e.g.,][]{isobe23b}. \citet{arellanocordova22b} reported Ne/O abundance ratios for three galaxies at $z> 7$ using spectra from the \textit{JWST} Early Release Observations \citep[ERO,][]{pontoppidan:22}. Comparing their results with local SFGs, the authors found that Ne/O does not appear to evolve with redshift.
In the recent work by \citet{isobe23a}, they studied the Ne/O ratio for a large sample of SFGs at $z\sim4-10$ in comparison with chemical evolution models \citep{watanabe23}. These authors also found a comparable results to those observed in galaxies at $z\sim0$ \citep[e.g.,]{izotov06, isobe22, arellanocordova22b}.

Unfortunately, being unable to estimate \Te\ in most of the high-$z$ galaxies in \citet{isobe23a} makes it hard to interpret the redshift evolution of Ne, S, and Ar. In general, accurate determinations of chemical abundances (e.g., C, N, Ne, S, Cl, Ar, and Fe) relies on two key aspects (1) the robust measurement of \Te~\citep[and electron density, see also,][]{sanders16, isobe23b, mendez-delgado23b} and (2) accurate ionization correction factors (ICFs, essential to account for the unseen ions) \citep[e.g.,][]{peimbert69, thuan95, izotov06, perez-montero07, dors16, peimbert17, berg21, amayo21}.

The nebular structures are complex in nature, the issue here is that the integrated observations do not capture this complexity. In principle, ICFs are constructed under the prescription of a single ionization source or \hii\ region, resulting in a potentially large bias when these are applied to the integrated spectra of SFGs. This could result in spurious trends that are not compatible with the prediction of chemical evolution models \citep[e.g.,][]{alexander23, watanabe23}, and in turn, bias our interpretation of the abundance patterns in the early Universe. 
Therefore, to decipher the chemical evolution of these elements, it is more important than ever to establish samples of nearby galaxies that span an ample range of conditions in order to correctly understand that physical properties of galaxies at high-$z$.
Local SFGs allow us to disentangle the ionization and temperature structure \citep{guseva12, izotov06, izotov12, mingozzi22, arellanocordova22b}. Thus, they provide a robust path to generate tools to understand the abundance ratios of metals from $z\sim0$ to $z>4$ in galaxies already observed with the \textit{JWST} such as CECILIA \citep{strom23}, and enable us to be prepared for the upcoming observations of the \textit{Extremely Large Telescopes} (ELTs).

The main aim of this paper is to study the chemical evolution of Ne, S, Cl, and Ar \footnote{N/O and C/O abundances will be presented in upcoming CLASSY papers}. These elements are synthesized in the interiors of massive stars as O. Therefore, their study can help us to constrain chemical evolution models of galaxies. Particularly, to compare observations of SFGs at different redshifts with theoretical models to understand the role of the CCSNe yields, which still suffer from large uncertainties. In this context, we use a sample of local SFGs from the COS Legacy Archive Spectroscopic SurveY \citep[CLASSY,][]{Berg22, james22} with properties similar to high-$z$ galaxies (stellar mass, star-formation rate, and ionization parameter) characterizing robustly their physical conditions and ionic abundances. With CLASSY, we can inspect the ICFs to provide information concerning the potential biases affecting the computation of total abundance ratios. The analysis of the CLASSY galaxies will be a robust template to interpret the chemical evolution of metals across cosmic time.  

The structure of this paper is as follows: In Section~\ref{sec:sample} we present the sample used in this analysis. Section~\ref{sec:physical conditions} describes the methodology utilized to derive the physical condition and ionic abundances. The analysis of the ICFs and the total abundances is also described. The results for the abundance patterns of CLASSY are presented in Section~\ref{sec:alpha}. In Section~\ref{sec:high-z}, we compare our results with the abundance patterns of high-$z$ galaxies and the bias in abundance determinations at high-$z$. In Section~\ref{sec:scaling-relations}
we also discuss the evolution of some scaling relations. In Section~\ref{sec:conclusion}, we summarize our conclusions. 

In this paper, we use the following solar abundance ratios taken from \citet{asplund21}: Ne/O$_{\odot} = -0.63\pm0.06$, S/O$_{\odot} = -1.57\pm0.05$,  Cl/O$_{\odot} = -3.38\pm0.20$, and  Ar/O$_{\odot} = -2.31\pm0.11$ with a solar oxygen abundance of 12+log(O/H) = 8.69$\pm$0.04. 

\section{Sample}\label{sec:sample}
\subsection{Main sample}
\begin{figure*}
\begin{center}
    \includegraphics[width=0.95\textwidth, trim=0 0 0 0,  clip=yes]{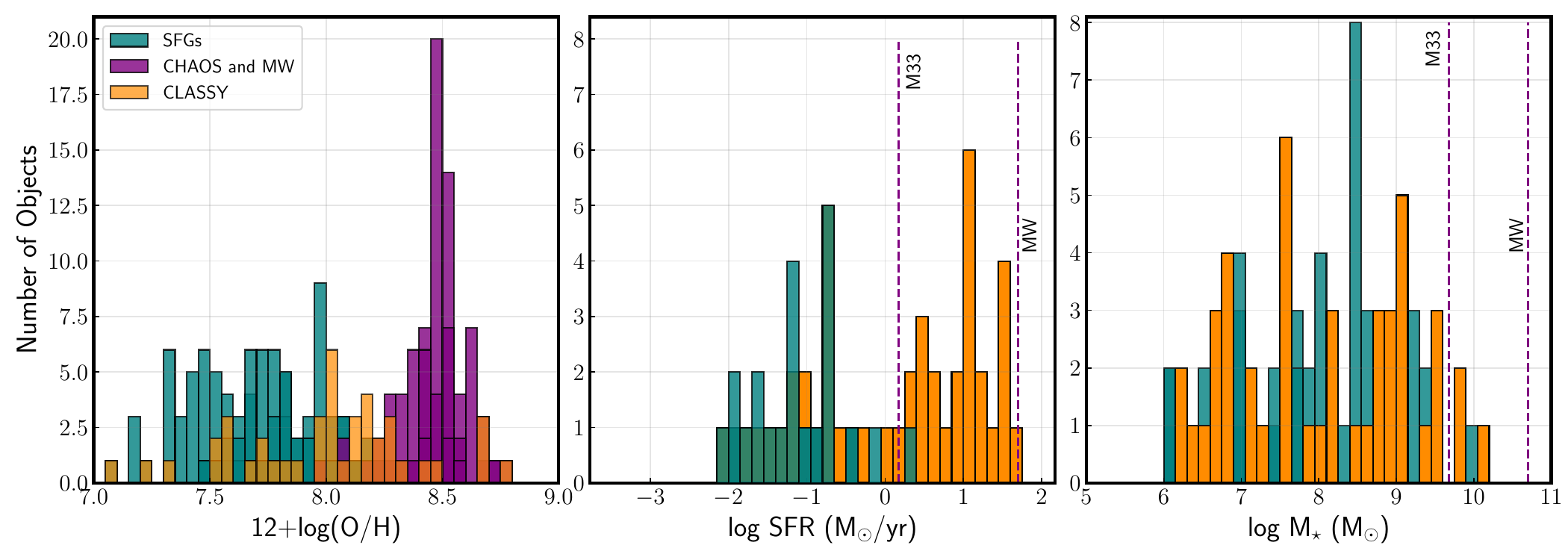}
        \caption{Metallicity, SFR and stellar mass of the CLASSY sample in comparison to the additional sample of SFGs and \hii\ regions collected from literature \citep{berg12, berg16, berg19a, izotov12, izotov17, croxall16, arellano-cordova2020b, arellanocordova21, Rogers22}. The dashed lines identify the range in SFR and stellar mass of M33, and the Milky Way (MW) taken from \citet{molinari22}, \citet{williams18}, \citet{berg20} and \citet{Licquia15}, respectively.}
\label{fig:hist_sample}
\end{center}
\end{figure*}

We use the ancillary optical spectra of 43 local ($0.002<$ $z$ $<0.18$) SFGs from CLASSY~\citep[][]{Berg22, james22}. This collection of optical spectra comprises observations from the APO/SDSS, LBT/MODS, VLT/MUSE/VIMOS and MMT telescopes/instruments, with a broad range of physical properties (see Figure~\ref{fig:hist_sample}) such as reddening ($0.02<$ $E(B-V)$ $<0.67$),  electron density, ($10<n_e$ (cm$^{-3}) <1120$), metallicity ($7.0<$ 12+log(O/H) $<8.7$), a proxy of the ionization degree ($0.5<$ [\ion{O}{3}] \W5007/[\ion{O}{2}] \W3727 $<38.0$), 
stellar mass ($6.2<$ log ($M_\star$/$M_\odot$) $<10.1$), and star formation rate ($-2.0<$ log SFR ($M_\odot$ yr$^{-1}$)  $<+1.6$). The details concerning the optical observation are reported in \citet[hereafter \PI]{Berg22}, \citet[][hereafter \PIV]{mingozzi22}, and \citet[][hereafter \PV ]{arellanocordova22a}.

\subsection{Comparison samples}
In addition, we have compiled different samples of SFGs and \ion{H}{2} regions of disc galaxies to compare with the results of the abundance patterns of CLASSY. For this additional sample it was also possible to derive the chemical abundances using the measurements of electron temperature. 

\underline{Integrated SFGs:} We selected a sample of local dwarf SFGs reported in \citet{berg12}, \citet{berg16},  \citet{izotov06},\citet{izotov12}, \citet{izotov17} and \citet{berg19}. We selected this sample because it covers lower-metallicities (7.0 $<$ 12+log(O/H) $< 8.2$), and the abundances are based on the \Te-sensitive method. Therefore, this sample will be useful to compare with our results for CLASSY.

\underline{\hii\ regions:} We have selected  44 Galactic \hii\ regions of \citet{arellano-cordova2020b} and \citet{arellanocordova21} with measurements of C, N, O, S, Cl, and Ar. This sample of Galactic \hii\ regions covers a range in metallicity of 8.0 $<$ 12+log(O/H) $<$ 8.9. In addition, we use \hii\ regions from 
The Chemical Abundances Of Spirals (CHAOS) project, which comprises $\sim$ 200 \hii~regions with chemical abundance determinations using the robust measurements of electron temperature \citep{croxall16, berg20, Rogers21, Rogers22}. We have selected those \hii\ regions reported in M33 \citep{ Rogers22}, which cover a range in metallicity between 8.0 $<$ 12+log(O/H) $<$ 8.7. The chemical abundances of CHAOS were derived following a methodology similar to this work as we explain below. 


\section{Nebular Physical Conditions}

\label{sec:physical conditions}

To analyze the nebular gas of the CLASSY sample, we utilized the nebular analysis package PyNeb \citep[][version 1.1.14]{luridiana15} to calculate the electron density (\Ne) and (\Te) and chemical abundances.  The physical conditions are based on our selection of atomic data using the 5-level atom model \citep{derobertis87}. The atomic data used in this study are listed in Table~\ref{tab:atomic_data}.

\input{atomic_data} %

We determine the physical conditions and chemical abundances for all samples using a uniform method consistent with the procedure utilized for CLASSY. We use the reddening corrected fluxes compiled from SFGs reported in \citet{berg16, berg19a, izotov06, izotov17, izotov21d}. While for the sample of \ion{H}{2} regions (CHAOS and the Milky Way \hii\ regions), we have taken the chemical abundances reported in the original papers. In Figure~\ref{fig:hist_sample}, we show the range in metallicity (left), SFR (middle), and stellar mass (right) covered by the CLASSY sample and the additional sample. 

\Te\, and \Ne\, are key to derive accurate chemical abundances. In particular, it requires the detection of faint emission features sensitive to the electron temperature such as the [\ion{O}{3}] \W4363, 
[\ion{N}{2}] \W5755, 
[\ion{S}{3}] \W6312, and
[\ion{O}{2}] \W\W7320,7330 auroral lines. 
Fortunately, most of the CLASSY sample we detect at least one $T_{\rm e}$-sensitive line and for others it is possible to trace the temperature structure using $T_{\rm e}$-diagnostics (see \PI\ and \PIV). In \PI\ we have also identified those \Te\-diagnostic used in the determination of the metallicity.  
For those galaxies with higher metallicity, we also inspected the [\ion{O}{3}] \W4363 profile to avoid any contamination of [\ion{Fe}{2}]\W4359 \citep[][ 12$+$log(O/H)~$>8.4$]{curti+17} to the observed flux of [\ion{O}{3}] \W4363, implying an underestimate of \TO\, \citep[see also \PIV ;][]{arellano-cordova20}. 


\subsection{ $n_e$ Measurements}


\Ne\ was calculated for the CLASSY galaxies using the [\ion{S}{2}]~\W6731/\W6717 diagnostic ratio, which primarily traces the low ionization region of the gas. A detailed analysis of different density diagnostics is presented in \PIV\, which shows no significant differences in the metallicities calculated using other density diagnostics. Abundances are insensitive to densities below \Ne $\sim$ 5000 cm$^{-3}$ for 12+log(O/H) ($<$ 0.2 dex difference than the low density limit), and are insensitive to even higher densities for other ratios like Ne/O and Ar/O \citep{osterbrock06}.

Density uncertainties were determined using a Boot Strap Monte Carlo error analysis. 
For each galaxy we generated a 1000 mock densities calculated by sampling [\ion{S}{2}] 
ratios from a Gaussian distribution centered at the observed intensity ratio and
with a width equal to the ratio uncertainty.
The standard deviation of the 1000 simulated densities was then taken as the 
uncertainty on the calculated density.


\subsection{ $T_e$ Measurements}

We compute the electron temperature for the CLASSY galaxies with the following $T_{\rm e}$-diagnostic ratios: [\ion{O}{2}]~\W\W3726,3729/(\W\W7319,7320+\W\W7330,7331)\footnote{Hereafter referred to as [\ion{O}{2}]~\W3727 and [\ion{O}{2}]~\W\W7320,7330 since such lines are blended due to the spectral resolution of the sample.}, [\ion{N}{2}] \W\W6548,6584/\W5755, [\ion{S}{3}] \W\W9069,9532/\W6312, and [\ion{O}{3}] \W\W4959,5007/\W4363, for the low-\To\ (or -\TN), intermediate-\TS, and high-\TO\ ionization temperatures, respectively. It is important to mention that the [\ion{S}{3}] lines can be strongly contaminated by telluric absorption in ground-based spectra, which is difficult to assess without the detection of [\ion{S}{3}]~\W9532, which is typically more strongly affected \citep{vacca03}, specially in $z =0$ objects.
For most of the spectra, it was only possible to measure [\ion{S}{3}]~\W9069. We have used it with the [\ion{S}{3}] theoretical emissivity ratio to estimate the intensity of [\ion{S}{3}]~\W9532 ([\ion{S}{3}] \W9532/\W9069 = 2.47).  In Table~\ref{tab:physical_conditions} in Appendix \ref{appen:tables}, we list the results for $n_{\rm e}$ and $T_{\rm e}$ associated with the different diagnostics used in columns 2-6. These temperatures represent different ionization regions of the galaxy (see also \PIV). 

Only one or two \Te-diagnostics were available for each galaxy. These diagnostics trace different regions of the gas based on the ionization structure of the nebula. 
If only one of \Te-diagnostics is available, \Te--\Te\ relations are used to estimate the unavailable \Te\ for a specific ionization region. To determined the complete temperature structure of each region, we have assessed some temperature relations from the literature \citet[e.g.,]{garnett90, izotov06} using CLASSY. We inspected such temperature relations minimize the uncertainties on the estimate of the missing temperature. in Appendix~\ref{appen:temperatures}, we compare several theoretical and empirical \Te\ - \Te\ relationships and find that the relationships of \citet{garnett92} work well to estimate \tlow and \tint from \TO. Additionally, we adopt the relationships of \citet{Rogers21} when \TO\ is not measured.
In Table~\ref{tab:physical_conditions}, we present the results for \tlow, \tint, and \thigh\ in columns 7-9, which represents the temperature structure for the CLASSY sample in this paper.  

To calculate the uncertainties associated with the electron temperatures and density, we use Monte Carlo simulations. We generated a Gaussian distribution of 1000 random values for each line intensity of each galaxy. The distribution was centered at the observed intensity, with a sigma equal to the associated uncertainty derived using standard deviation.

\section{Nebular Abundances}

We use the following equation to calculate ionic abundances relative to hydrogen:
\begin{equation}
    \frac{N(X^i)}{N(H^+)} = \frac{I_{\lambda(X^i)}}{I_{{\rm H}\beta}}
                            \frac{j_{{\rm H}\beta}}{j_{\lambda(X^i)}},  
\end{equation}
where $j_{\lambda(X^i)}$, the emissivity coefficient, is sensitive to the 
adopted temperature.
We derive the ionic abundances of O, Ne, S, Cl, and Ar, using the three-zone temperature structure adopted in Section~\ref{sec:physical conditions} 
($T_{\rm e}$(Low), $T_{\rm e}$(Int), and $T_{\rm e}$(High)). 
We use $T_{\rm e}$(Low) to calculate the ionic abundances of the low ionization ions such as O$^{+}$ and S$^{+}$. For O$^{+}$, we use the measurements of [\ion{O}{2}] \W3727 when available and [\ion{O}{2}] \W\W7320,7330 lines for the rest of the galaxies. For S$^+$, we use the measurements of  [\ion{S}{2}] \W\W6717,6731 lines, respectively.
To calculate the ionic abundances of intermediate ionization ions, S$^{2+}$,  Cl$^{2+}$, and Ar$^{2+}$, we use the measurements of  [\ion{S}{3}] \W6312, 9069, [\ion{Cl}{3}] \W5518, 5538,  and [\ion{Ar}{3}] \W7135, 7751 lines, and $T_{\rm e}$(Int) as a representative temperature of the intermediate ionization zone. For S, we have determined the S$^{2+}$ using [\ion{S}{3}] \W6312, which allows us to increase the number of galaxies with measurements of S$^{2+}$. Therefore, for those galaxies we measurements of both [\ion{S}{3}] \W6312 and [\ion{S}{3}] \W9069, we calculated consistent results of the ionic abundance of S$^{2+}$, which provide for a reliable use of [\ion{S}{3}] \W6312. For Cl, we determined the abundance only when both lines are measured to ensure an accurate determination of Cl$^{2+}$. The ratio of [\ion{Cl}{3}] \W5518, 5538 is also a density diagnostic (see also \PIV).
Finally, we used $T_{\rm e}$(High) to derive the ionic abundances for high and very high ionization species, O$^{2+}$, Ne$^{2+}$, Cl$^{3+}$ and Ar$^{3+}$, using the [\ion{O}{3}] \W5007, [\ion{Ne}{3}] \W3869, [\ion{Cl}{4}] \W8049,  [\ion{Ar}{4}] \W\W4711,41 lines, respectively.   

\begin{deluxetable}{l c c  c c } 
\tablecaption{Ionic abundances of the CLASSY galaxies with detection of both [\ion{Cl}{3}] \W\W5518,38 and [\ion{Cl}{4}] \W8049 in CLASSY}
\label{tab:chlorine} 
\tablewidth{0pt}
\tablehead{
CLASSY   &  [\ion{Cl}{4}] \W 8049 &$^{a}$Cl$^{2+}$    &   $^b$Cl$^{3+}$  & O$_{\rm 32}$}
\startdata
J0944-0038 & 0.19 $\pm$ 0.01 & \nodata  & 3.76 & 8.2  \\
J1044+0353 & 0.22 $\pm$ 0.03 &  3.18   &  3.73 & 3.6\\
J1150+1501 & 0.09 $\pm$ 0.02 &  4.40   & 3.65  &  2.9\\
J1225+6109 & 0.12 $\pm$ 0.03 &  4.08   & 3.65  &  6.8 \\
J1253-0312 & 0.21 $\pm$ 0.02 &  4.22   & 3.90  & 7.5\\
J1323-0132 & 0.25 $\pm$ 0.01 &  \nodata & 3.77 & 38.0 \\
J1448-0110 & 0.16 $\pm$ 0.03 &  4.13     & 3.82 & 7.3\\
J1545+0858 & 0.22 $\pm$ 0.01 & \nodata  & 3.78 & 9.2\\
\enddata
\tablecomments{$^a$In units of 12+log(Cl$^{2+}$/H$^+$). 
$^b$ In units of 12+log(Cl$^{3+}$/H$^+$). The ionization parameter is defined as O$_{32}$ = [\ion{O}{3}] \W5007/[\ion{O}{2}] \W3727.} 
\end{deluxetable}

\subsection{ICF Tests}
The significant line detections from O, Ne, S, Cl, and Ar in the CLASSY sample 
allow us to calculate their total abundances.
For oxygen, the total O/H abundances were calculated in \citet{Berg22} as the sum 
of the O$^+$/H$^+$ and O$^{2+}$/H$^+$ ion fractions, as the emission from O$^0$ and 
O$^{3+}$ is negligible in typical star-forming regions \citep[e.g.,][]{berg21}.
The total O/H abundances are reproduced in Table~\ref{tab:abundance_ratios} in 
Appendix~\ref{appen:tables}.

On the other hand, unlike oxygen, not all relevant ionic species are observed for the other 
$\alpha$-elements and so it is necessary to use an ionization correction factor (ICF) to 
account for the unobserved ionic species. 
To ensure robust total abundance determinations of these elements, 
we explored the performance of different ICFs.

First, we examine the ICFs of \citet{amayo21}, which are focused on C, N, Ne S, Cl, and Ar. 
These ICFs are based on a grid of photoionization models from the Mexican Million Models 
database \citep[][]{morisset09}, cover the physical properties of both \ion{H}{2} regions
and integrated low-mass galaxies. 
covering a broad range of physical properties of 
extragalactic \ion{H}{2} regions and blue compact dwarf galaxies (BCDs). 
The parametric expressions of these ICFs depend on the ionization parameter, 
measured as O$^{2+}$/(O$^{+}$ + O$^{2+}$). 

Previous studies focused on the $\alpha$-elements in SFGs have proposed different sets of ICFs.
In this context, we have analyzed the ICFs proposed by \citet{izotov06}, 
which are based on photoionization models of N, Ne, O, S, Cl, Ar, and Fe by \citet{stasinska03}. 
These ICFs also depend on the ionization parameter, but are calibrated for three different 
ranges of metallicity of integrated SFGs. 
For Ne, S and Ar, we have also explored the ICFs of \citet{thuan95},
which are based on models covering a the properties in only local integrated SFGs,
while and \citet{dors13} is based on \ion{H}{2} regions and SFGs.
The ICFs of both \citet{thuan95} and \citet{dors13} were calibrated on samples including 
both \hii\ regions and integrated galaxies.
Finally, for Cl, we include a new empirical ICF based on \hii\, regions with high spectral 
resolution from \citet{Dominguez-Guzman2022}.

The
ICFs derived by the different authors depend on the assumptions made for the nebular 
and stellar properties of the \hii\ regions. 
For example, different input stellar spectral energy distributions (SED) parameters have been 
adopted for different studies. 
Major ingredients are the stellar evolution and atmosphere models. 
\citet{izotov06} and \citet{dors13} can be compared directly as their ICFs are based on the 
same set of stellar models from Starburst99 \citet{leitherer99}. 
Note that \citet{izotov06} used the photoionization models of \citet{stasinska03} but with 
the more modern Starburst99 stellar models instead of a previous model generation. 
\citet{amayo21} adopted stellar SEDs obtained with the PopStar code \citet{molla09}, 
whose stellar atmosphere models for massive hot stars are identical to those in Starburst99. 
However, the isochrones in PopStars are based on Padova evolution models, 
as opposed to the Geneva models in Starburst99. 
 Figure 4 of \citet{vazquez&leitherer05} highlights the significant differences 
between SEDs computed with Geneva and Padova tracks. 
The earlier ICFs of \citet{thuan05} were obtained with an unevolved zero-age main-sequence population of hot stars using unblanketed atmosphere models of \citet{mihalas72}. 
Therefore these SEDs produce a much harder ionizing spectrum than those discussed before.

Below we calculate the ionic abundances of Ne, Ar, S, and Cl and use them to
analyze the ICFs using the 
Ne$^{2+}$/O$^{2+}$, Ar$^{2+}$ /O$^{2+}$ and (Ar$^{2+}$ + Ar$^{3+}$)/O$^{2+}$ , 
(S$^{+}$ + S$^{2+}$)/(O$^{+}$ + O$^{2+}$), and Cl$^{2+}$/O$^{2+}$ ionic ratios, 
respectively.
To test the ICFs, we compare to the observed CLASSY galaxy trends.
We have used the solar ratio of Ne/O$_{\odot}$, and S/O$_{\odot}$ for scaling the ICFs 
with respect to the ionic abundance ratio used in the computation of the total abundance 
(e.g., Ne$^{2+}$/O$^{2+}$). 
Therefore, those ICFs with similar trends to those used in the CLASSY galaxies analyzes 
should result in total abundances close to the solar abundance. 
We also examined the resulting total abundance trends of these elements as a second test
of the ICFs. 
In this context, we have analyzed different ICFs to find those most suitable the CLASSY sample 
and for SFGs with similar properties to the CLASSY galaxies. 

\subsection{Ionic Abundances and Relative $\alpha$-Abundances}
\label{sec:alpha}
In principle, Ne, S, Cl, and Ar production depends very little on metallicity 
because these elements are mainly synthesized in massive stars by the 
$\alpha$-capture process.  
Therefore, any dependence on metallicity of the abundance ratio relative 
to O might be associated with other properties such as the degree of ionization, 
dust depletion, the ICF, and/or observational problems \citep[e.g.,][]{amayo21}. 
We start by investigating the ICFs below.


\begin{figure}[h]
\begin{center}
    \includegraphics[width=0.45\textwidth, trim=0 0 0 0,  clip=yes]{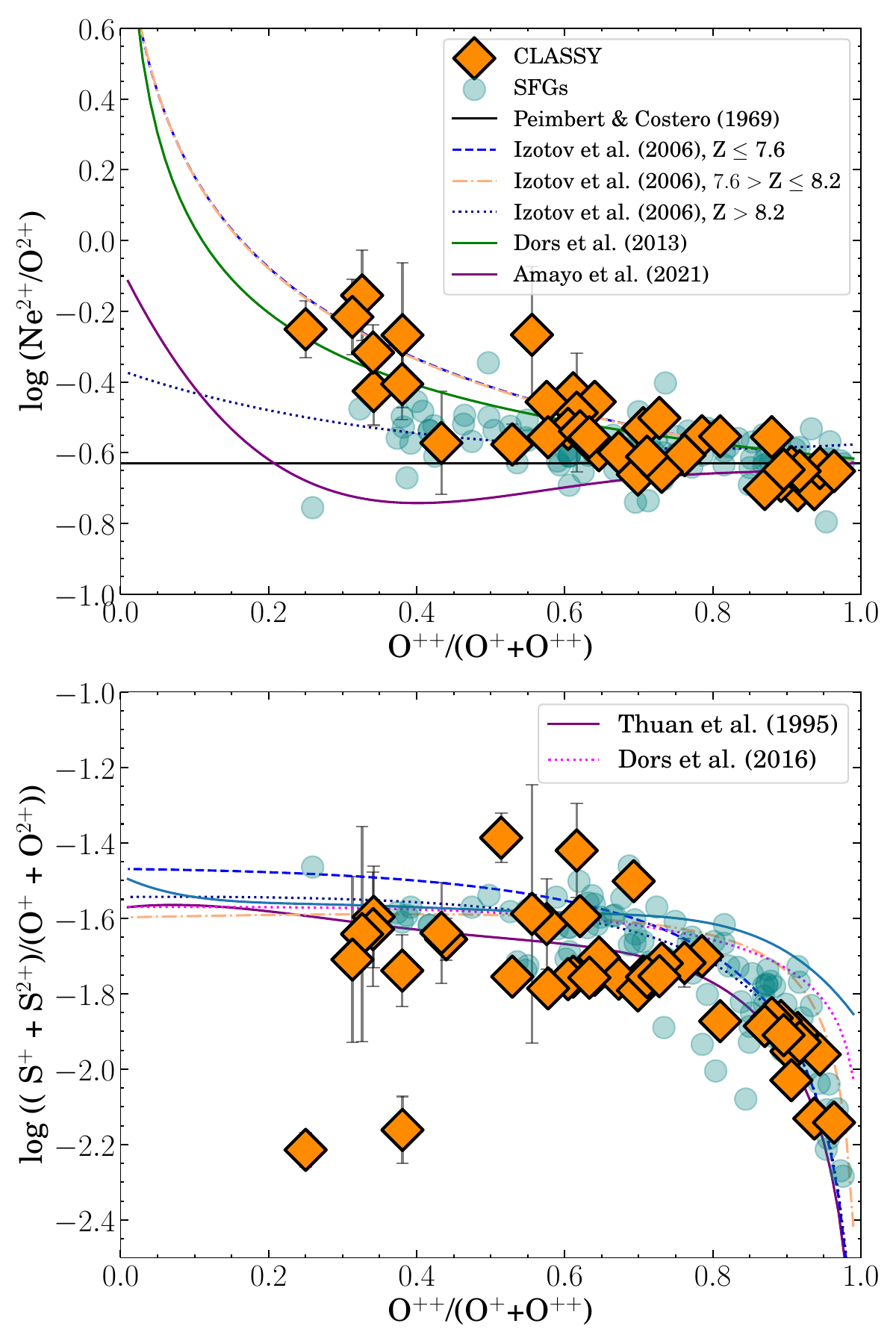}
        \caption{ Ionic abundance ratios of Ne, S, and O used to correct for the unseen ionic species. The different lines represent the ICFs derived by \citet{peimbert69}, \citet{thuan95}, \citet{izotov06}, \citet{dors13}, \citet{dors16}, and \citet{amayo21}. \textit{Top:} The ratio of Ne$^{2+}$/O$^{2+}$ as a function of the ionization degree measured as O$^{2+}$/(O$^{+}$ + O$^{2+}$) for the CLASSY galaxies with measurements of [\ion{Ne}{3}] \W3869. \textit{Bottom:} the ratio of (S$^{+}$ + S$^{2+}$)/(O$^{+}$ + O$^{2+}$) as a function of the ionization degree. The SFGs compiled from the literature are shown in circles. CLASSY galaxies follow the trends given by the ICFs of \citet{dors13} for Ne and \citet{izotov06} for S. `Z' indicates the value of 12+log(O/H). }
\label{fig:ICFS_test}
\end{center}
\end{figure}


\subsubsection{Neon}
\label{sec:neon}
For Ne, only one ionization state is observed, Ne$^{2+}$ (41.0–63.5 eV),
such that Ne/O abundances are typically determined using
\begin{equation}
    \frac{\rm Ne}{\rm O} = \frac{{\rm Ne}^{2+}}{{\rm O}^{2+}} \times {\rm ICF(Ne)},
\end{equation}
where Ne$^{2+}$ is determined from the [\ion{Ne}{3}] \W3868 emission line.
Note that a similar equation is used to determine Ne/H such that 
Ne/H = Ne$^{2+}$/H$^+ \times$ICF(Ne). 
We use the [\ion{Ne}{3}] \W3868 detection in 38 CLASSY galaxies 
(see Table~\ref{tab:abundance_ratios}) to calculate Ne$^{2+}$ and, subsequently, Ne/O.

We plot the Ne$^{2+}$/O$^{2+}$ ratio as a function of the ionization parameter 
(O$^{2+}/$(O$^{+}$ + O$^{2+}$)) in the top panel of Figure~\ref{fig:ICFS_test}.
The different lines illustrate the ICFs analyzed for Ne from 
\citet[][solid black]{peimbert69}, 
\citet[][blue dashed, orange dot-dashed, and navy dotted]{izotov06}, 
\citet[][solid green]{dors13}, and 
\citet[][solid purple]{amayo21}. 
Figure~\ref{fig:ICFS_test} shows that the Ne ICFs of 
\citet[][low and intermediate metallicity represented as "Z"]{izotov06} and \citet{dors13} 
have similar shapes as the Ne$^{2+}$/O$^{2+}$ trend for the CLASSY galaxies, 
while the ICF of \citet{amayo21} overestimates the Ne ICF, leading to an overestimate of 
the total Ne/O ratio with respect to the solar value.  

\begin{figure*}
\begin{center}
    \includegraphics[width=0.80\textwidth, trim=30 0 30 0,  clip=yes]{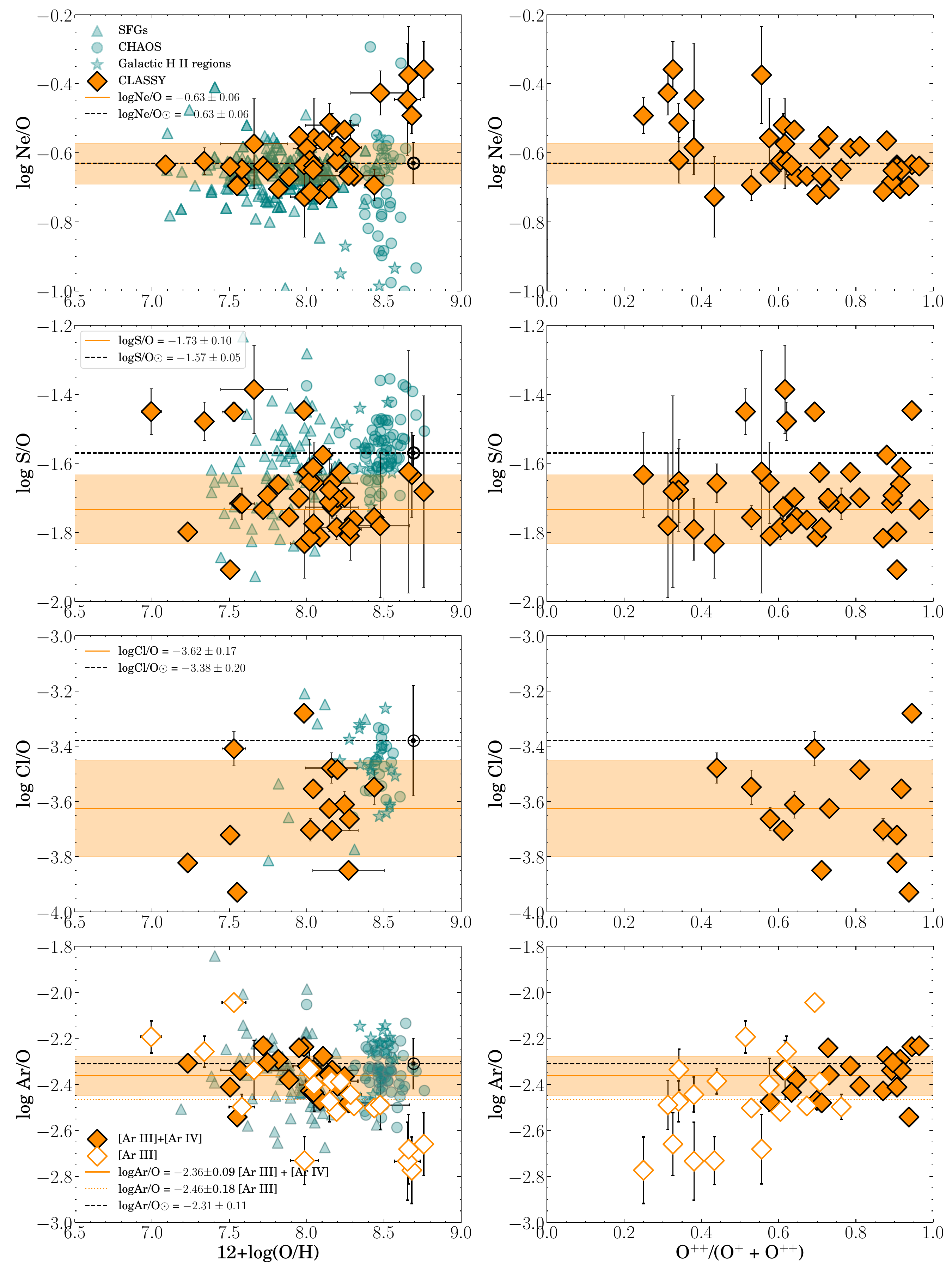}
        \caption{The Ne/O, S/O, Cl/O, and Ar/O ratios versus metallicity (left) and ionization degree (right) relationships for the CLASSY galaxies. In principle, the ICFs are corrected by ionization effects. The additional SFGs are from \citet{berg12, berg16, berg19a, izotov06, izotov12, izotov17}, \ion{H}{2} regions from the CHAOS survey \citep{Rogers21} of M33, and Galactic \ion{H}{2} regions from \citet{arellano-cordova2020b, arellanocordova21}. For Cl, we have also added three galaxies of \citet{izotov21} with measurements of [\ion{Cl}{3}] \W\W 5518, 5531 for comparison. The black solid symbol and dashed line represents the solar abundance of \citet{asplund21}. 
        The mean and dispersion values of the abundance ratios are illustrated with a solid line and shadow box color-coded for the CLASSY galaxies. For Ar/O, we present two different results depending on which emission lines are used to determine the total abundance. The filled symbols represent the value of Ar/O derived using [\ion{Ar}{3}] + [\ion{Ar}{4}], while that empty symbols illustrate the value of Ar/O derived using [\ion{Ar}{3}]. For Ar/O, the dashed and solid lines indicate the mean and standard deviation of solid and empty symbols, respectively.}
\label{fig:alpha_elements}
\end{center}
\end{figure*}

We present the Ne/O \textit{vs.}\ O/H relation in the top left panel of 
Figure~\ref{fig:alpha_elements}. 
The comparison samples are also plotted in the Ne/O--O/H plot, showing good agreement with the 
CLASSY sample for 12+log(O/H)$<8.4$
Interestingly, the \ion{H}{2} region samples have a large dispersion to low Ne/O abundances
at 12+log(O/H)$>8.0$.
The Ne/O ratio of the CLASSY galaxies follows the expected constant abundance pattern with an 
average value of log(Ne/O) = $-0.63\pm0.06$, in excellent agreement with the solar abundance 
reported by \citet[][log(Ne/O)$_{\odot} = -0.63\pm0.06$]{asplund21}. 
However, for five CLASSY galaxies, we have determined super-solar values of log(Ne/O) $> -0.4$ 
(J0808+3948, J1144+4012, J1521+0759, J1525+0757, and J1612+081), which is independent of the ICF 
of Ne involved in this study. 

The top right panel of Figure~\ref{fig:alpha_elements} shows the Ne/O ratio as a function of  
ionization (O$^{2+}$/(O$^{+}$ + O$^{2+}$)). 
It is clear that for low-ionization regions (high-metallicity), there is a large dispersion
extending to high values of Ne/O.
If we exclude these outliers, the relation between the ionization degree and Ne/O is constant 
as is expected. 
We also used this plot to inspect the 
applicable range of each ICF used to derive Ne/O.
In principle, the applicability of the Ne ICF of \citet{dors13} is restricted to values of 
O$^{2+}$/(O$^{+}$ + O$^{2+}$) > 0.2. 
This might explain the high values of Ne/O for the 4/5 galaxies with an ionization degree
below this limit (excludes J1521+0759 with O$^{2+}$/(O$^{+}$ + O$^{2+}$) = 0.56). 
However, note that the value of O$^{2+}$/(O$^{+}$ + O$^{2+}$) is highly sensitive to the 
temperature structure adopted. 
The five objects with high values of Ne/O are metal-rich, implying a decrease of the 
\Te-sensitive line [\ion{O}{3}] \W4363. 
Therefore, here we find Ne/O increases with metallicity, 
which has been attributed to the depletion of O into dust grains \citep{izotov06, guseva11}. 

We calculated the dispersion in Ne/O resulting from the use of these ICFS. 
For the ICFs of \citet{izotov06} and \citet{dors13}, we calculated similar dispersions of 
0.06 dex and 0.07 dex, respectively. 
The ICF of \citet{amayo21} produces a larger dispersion value of 0.12 dex and a strong 
correlation with metallicity. 
Therefore, we used and recommend the ICF of \citet{dors13} to derive the Ne/O ratios of 
CLASSY-like galaxies. 
We report the mean and dispersion values of Ne/O for the CLASSY sample in 
Table~\ref{tab:mean_dispersion}.


\subsubsection{Sulfur}
\label{sec:sulfur}
For sulfur, both S$^{2+}$ (22.3--34.8 eV) and S$^+$ (10.4–-22.3 eV) span 
the low-ionization zone, although S$^+$ emission can also extend into the neutral 
H zone. 
We do not correct for this issue, but the generally high-ionization of the CLASSY
sample suggest that that the S$^+$ contribution from outside the \ion{H}{2} region
should be minimal. 
On the high-ionization end, we must consider contributions from  S$^{3+}$ 
(34.8–-47.2 eV).
Thus, we determine the S/O abundance using
\begin{equation}
 \frac{\rm S}{\rm O} = \left(\frac{ {\rm S}^{+}  + {\rm S}^{2+}} {{\rm O}^+ + {\rm O}^{2+}}\right)
                          \times {\rm ICF(S)},
\end{equation}
where S$^{+}$ is determined from the [\ion{S}{2}] \W\W6717,6731 emission doublet and
S$^{2+}$ is determined from [\ion{S}{3}] \W\W6312,9069.
Note that a similar equation is used to determine S/H such that 
S/H = (S$^{+}$+S$^{2+}$)/H$^+ \times$ICF(S).

For the CLASSY S/O abundances, we only consider ICFs based on the use of both 
[\ion{S}{2}] and [\ion{S}{3}]. 
With this criterion, we used the resulting 41 CLASSY galaxies to calculate 
S$^{+}$, S$^{2+}$ and, subsequently, S/O.
In this analysis we used the ICFs of \citet{amayo21}, \citet{dors16}, and \citet{thuan95}.
Note that some ICFs only use one of the S ions. 
We plot the (S$^{+}$ + S$^{2+}$)/(O$^{+}$ + O$^{2+}$) ratio as a function of the ionization 
parameter (O$^{2+}/$(O$^{+}$ + O$^{2+}$)) in the bottom panel of Figure~\ref{fig:ICFS_test}.
With the exception of two low-ionization outliers, all three ICFs trace the trend of the 
CLASSY sample well. 
Using all ICFs, we calculated three sets of CLASSY S/O abundances.
All three S/O sets had dispersions around 0.15 dex, but the ICF of \citet{thuan95} provided
the closest values of S/O to solar (log(S/O = $-1.57$). 
Based on this analysis, we used and recommend the ICF provided by \citet{izotov06}.

The resulting CLASSY S/O abundances are plotted in Figure~\ref{fig:alpha_elements} 
as a function of metallicity (left column) and ionization parameter (right column).
The comparison samples are also plotted in the S/O--O/H plot, showing good agreement with the 
CLASSY sample for 12+log(O/H)$<8.4$
We also noticed in Figure~\ref{fig:alpha_elements} that the sample of SFGs of \citet{izotov06} shows a large dispersion but with most of the sample centered on the solar abundance. 
Note also the large uncertainties of the metal-rich galaxies. In a similar way, the sample of \ion{H}{2} regions shows a good agreement with the solar abundance in comparison with the CLASSY galaxies over the same range of metallicity. 

We note that five galaxies have S/O values that are higher than the main CLASSY sample of S/O: 
J0127$-$0619, J0405$-$3648, J0944+3442, 
J1132+5722, and J1253$-$0312. 
To confirm the high values of S/O, we inspect the ionic abundance of S$^{2+}$ derived using 
[\ion{S}{3}] \W6312 and [\ion{S}{3}] \W9069. 
For galaxies with both measurements, we find a consistent agreement between the value of 
S$^{2+}$ derived using  [\ion{S}{3}] \W6312 and [\ion{S}{3}] \W9069 
(i.e., J0405$-$3648 and J1132+5722). 
Further, we examined whether the adopted \Te\ affects the derived S/O for the five outliers. 
\citet{dominguezguzman19} and \citet{esteban20} show that changes in the temperature structure 
introduce slight variation to the S/O ratio (and Cl/O and Ar/O) increasing with metallicity, 
in particular when the two-zone temperature structure is used. 

In our analysis, for J0405$-$3648 and J1132+5722, we use the direct measurement of 
\TS\ (or \tint) to estimate S$^{2+}$, and for the rest of galaxies (J0127$-$0619, J0944+3442, 
and J1253$-$0312) with high S/O, we use the adopted temperature relations (see 
Section~\ref{sec:physical conditions}) to estimate \tint. 
We estimate the mean value between \tlow\ and \thigh\, whose result should be a proxy of \tint\ 
\citep[see also][]{dominguezguzman19}. 
We have used this procedure to preclude that the high value of S/O could be due to the $T_{\rm e}$ 
selected for those regions since \TS\ might be affected by telluric lines, a possible blend of 
[\ion{S}{3}]~$\lambda$6312 with [\ion{O}{1}]~$\lambda$6300, or reddening corrections that are 
biased low. 
We find that for these individual galaxies, the only two regions that show a significant 
change in the results are J0944$-$0038 and J1253$-$0312. 
Therefore, part of those differences might be associated with the estimate of \TS. 
In this context, we have excluded those five galaxies for our analysis of S. 
We determine a mean value of log(S/O) = $-1.73\pm$0.10 which is 0.16 dex lower than the 
solar value of log(S/O)$_\odot = -1.57\pm0.05$ using a number of 35 galaxies since we have 
discarded the five outliers galaxies.

\subsubsection{Chlorine}
\label{sec:chlorine}
The CLASSY sample provides the rare opportunity to trace the chemical abundances of Cl. 
The production of Cl is based on the radioactive decay of $^{37}$Ar formed by a 
single neutron capture of $^{36}$Ar \citep[see][]{clayton03, esteban15}.
For Cl, typically only one ionization state is observed, Cl$^{2+}$ (23.8–39.6 eV),
such that Cl/O abundances are typically determined using
\begin{equation}
    \frac{\rm Cl}{\rm O} = \frac{{\rm Cl}^{2+}}{{\rm O}^{2+}} \times {\rm ICF(Cl)},
\end{equation}
where Cl$^{2+}$ is measured via the [\ion{Cl}{3}] \W\W 5518,5538 doublet (see \PIV). 
Note that a similar equation is used to determine S/H such that 
Cl/H = Cl$^{+}$/H$^+ \times$ICF(Cl). 
For our analysis of Cl, we use a sample of 15 CLASSY galaxies with a significant 
detections of [\ion{Cl}{3}] to determine the Cl$^{2+}$ abundance and, subsequently, Cl/O.
\footnote{Note that [\ion{Cl}{3}] \W\W5518,5538 detections were not possible in the 
LBT/MODS spectra of three CLASSY galaxies (J0944$-$0038, J1323$-$0132, and J1545+0858)
due to the wavelength gap between the blue and red arms.
\citep[see Figure 2 of][]{arellanocordova22a}.}

While [\ion{Cl}{3}] \W5518,5538 are the brightest Cl lines in SFGs,
and contributions from other ionization states of Cl are typically negligible 
\citep[e.g.,][]{esteban15}, we detect Cl$^{3+}$ via [\ion{Cl}{4}] \W8049 in eight 
CLASSY galaxies (see Table~\ref{tab:chlorine}). 
With an ionization potential of $\sim$39.6 eV, this line is only detected
in high-ionization objects \citep[e.g.,][]{esteban15}.
We detect both Cl$^{2+}$ and Cl$^{3+}$ for five CLASSY galaxies,
enabling a more direct measure of chlorine using Cl = Cl$^{2+}$ + Cl$^{3+}$. 
However, Cl$^+$ has an ionization potential of 13.0--23.8 eV, similar to O$^+$,
and so an ICF should still be used to account for the potentially significant contribution
from Cl$^+$ \citep{esteban15, arellano-cordova2020b, amayo21}. 

We compared Cl ICFs from \citet{izotov06}, \citet{amayo21}, and \citet{Dominguez-Guzman2022}. 
We found that the ICF of \citet{izotov06} has the least dependence on metallicity and ionization 
parameter, and, thus, adopt the Cl ICF from \citet{izotov06} to derive Cl/O abundances for CLASSY.
 We show the resulting Cl/O abundances versus O/H in the third row of 
Figure~\ref{fig:alpha_elements}. 
On average, we find the CLASSY galaxies to have subsolar Cl/O abundances, 
with a mean value of log(Cl/O) $= -3.62\pm0.17$.
This value is 0.24 dex lower than the solar value,
but is consistent with previous studies of SFGs \citep[e.g.,][]{izotov06,izotov17}.
We note that similar trends are obtained using the ICFs of \citet{amayo21} and 
\citet{Dominguez-Guzman2022}. 

On the other hand, \citet{arellano-cordova2020b} used a sample of Galactic \hii\ regions 
to derive Cl/O abundances using the same ICFs as this work, but found a good agreement with 
log(Cl/O)$_\odot$ = $-$3.38$\pm$0.20 \citep{asplund21}.
Similarly, \citet{Rogers22} found consistent results with Cl/O$_\odot$ for the CHAOS sample 
of \ion{H}{2} regions in M33 using the ICF of \citet{amayo21}. 
However, the photospheric solar Cl abundance is somewhat uncertain,
making it difficult to assess the significance of the subsolar Cl/O abundances of CLASSY.
Nonetheless, it is interesting that the integrated S/O and Cl/O abundances of the CLASSY galaxies 
\citep[and other dwarf SFGs, e.g.,][]{berg12, berg16} are lower than the abundances of
individual \hii\ regions within galaxies \citep[e.g.,][]{izotov06}. 

The right column of Figure~\ref{fig:alpha_elements} shows the Cl/O ratio with respect to  
ionization parameter. 
We do not find a significant correlation of Cl/O with the ionization parameter, 
although there is a large dispersion in the data.
This dispersion may be due to observational uncertainties associated with the 
faintness of the [\ion{Cl}{3}] lines.
Higher S/N [\ion{Cl}{3}] and [\ion{Cl}{4}] observations are needed to determine 
how observational and ICF uncertainties are affecting the Cl/O--O/H trend.

\subsubsection{Argon}
\label{sec:argon}
For Ar, only the Ar$^{2+}$ ionization state is observed for the majority
of the CLASSY optical spectra via detections of [\ion{Ar}{3}] \W7135
(significantly detected in 43 galaxies).
Additionally, Ar$^{3+}$ (40.74--59.81 eV) is observed as [\ion{Ar}{4}] \W\W4711,4740\footnote{ 
Note that [\ion{Ar}{4}] \W4711 is often blended with the \ion{He}{1} \W4713 line in low- to
moderate-resolution spectra ($R \lesssim 2000$), and so care must be taken to account for the
\ion{He}{1} contribution to the flux. }
for 22 of these CLASSY galaxies such that Ar$^{2+}$ + Ar$^{3+}$ can be used for Ar 
abundance determinations.
However, the ionization potential of Ar$^+$ (15.76--27.63 eV) overlaps with the
low-ionization zone.
Therefore, an ICF must be used to account for the unseen Ar$^+$, and sometimes Ar$^{3+}$, 
ions in Ar abundance determinations using:
\begin{equation}
    \frac{\rm Ar}{\rm O} = \frac{{\rm Ar}^{2+} + {\rm Ar}^{3+}}{{\rm O}^{2+}} \times {\rm ICF(Ar)}.
\end{equation}
Note that a similar equation is used to determine Ar/H such that 
Ar/H = (Ar$^{2+}$+Ar$^{3+}$)/H$^+ \times$ICF(Ar).
Additionally, some works only use Ar$^{2+}$ in their ICFs and abundance calculations.

We present Ar abundances using two different approaches:
(i) using both \fariii\ \W7135 and \fariv\ \W\W 4711,4741 lines (S/N $\geq$ 3)
for a sample of 22 SFGs (the preferred method) and
(ii) using only \fariii\, \W7135 (S/N $\geq$ 3) for the remaining 20 galaxies with 
\fariii\, \W7135 detections. 
We test  different ICFs for Ar from \citet{thuan95}, \citet{izotov06}, and \citet{amayo21}. 
For method (i) using only \fariii, the ICF(Ar$^{2+}$) results in a strong dependence on 
metallicity for 12+log(O/H) $> 8.2$ and a large dispersion up to 0.16 dex. 
For method (ii) using both \fariii\ and \fariv, we find that the ICF(Ar$^{2+}$ + Ar$^{3+}$) 
of \citet{izotov06} and \citet{amayo21} are in excellent agreement, providing similar mean 
and dispersion values of log(Ar/O) = $-2.36\pm0.09$ and log(Ar/O) = $-2.35\pm0.09$, respectively. 
The ICF(Ar$^{2+}$ + Ar$^{3+}$) of \citet{thuan95} also provides similar results.  

In the bottom row of Figure~\ref{fig:alpha_elements} we present the Ar/O abundances for 
the CLASSY sample. 
The open diamonds represent those SFGs with only detections of \fariii\, 
while the solid diamonds represent those with detections of both \fariii\ and \fariv. 
For the latter subset, we also determined Ar/O using only the detection of \fariii\ and
found differences between the two methods of up to $\Delta\log({\rm Ar/O})\lesssim0.10$ dex, 
but with $\Delta\log({\rm Ar/O})\lesssim0.05$ dex for most galaxies. 
This implies that robust Ar/O determinations require detections of both \fariii\ and \fariv, 
particularly for high ionization objects \citep[see also,][]{berg21}. 
Therefore, we adopt the ICF(Ar$^{2+}$ + Ar$^{3+}$) of \citet{izotov06} 
(see also Table~\ref{tab:mean_dispersion}).

Using the more robust \fariii\ $+$ \fariv\ sample of CLASSY Ar/O abundances,
we find a mean value of log(Ar/O) $= -2.36\pm0.09$ that is in good agreement with both the 
solar abundance (log(Ar/O)$_\odot = -2.31\pm0.11$) and the literature samples at lower 
metallicities (12+log(O/H) $< 8.3$). 
If we include the \fariii-only sample, however, then the Ar/O vs. O/H relation shows an 
unexpected trend with metallicity at higher metallicities.
Other studies of integrated galaxy spectra also have reported a dependence with metallicity 
in the Ar/O--O/H relation \citep[e.g.,][]{perez-montero07, Kojima2021}. 
On the other hand, the literature sample of \ion{H}{2} regions is consistent with the solar 
Ar/O ratio even at higher metallicities. 

The method (i) Ar/O abundances also show a trend with O$^{2+}$/(O$^{+}$ + O$^{2+}$) in the right 
column of Figure~\ref{fig:alpha_elements} (open diamonds), which could be associated with the 
performance of the ICF with respect to ionization degree.
In comparison, the values of Ar/O for galaxies with measurements of [\ion{Ar}{4}] are constant 
for O$^{2+}$/(O$^{+}$ + O$^{2+}$) $>$ 0.6. 
Therefore, the performance of different ICFs may be more uncertain at high-metallicities and
low-ionization parameter \citep[e.g., ][]{amayo21}. 


\begin{figure*}[h]
\begin{center}
    \includegraphics[width=0.85\textwidth, trim=0 0 0 0,  clip=yes]{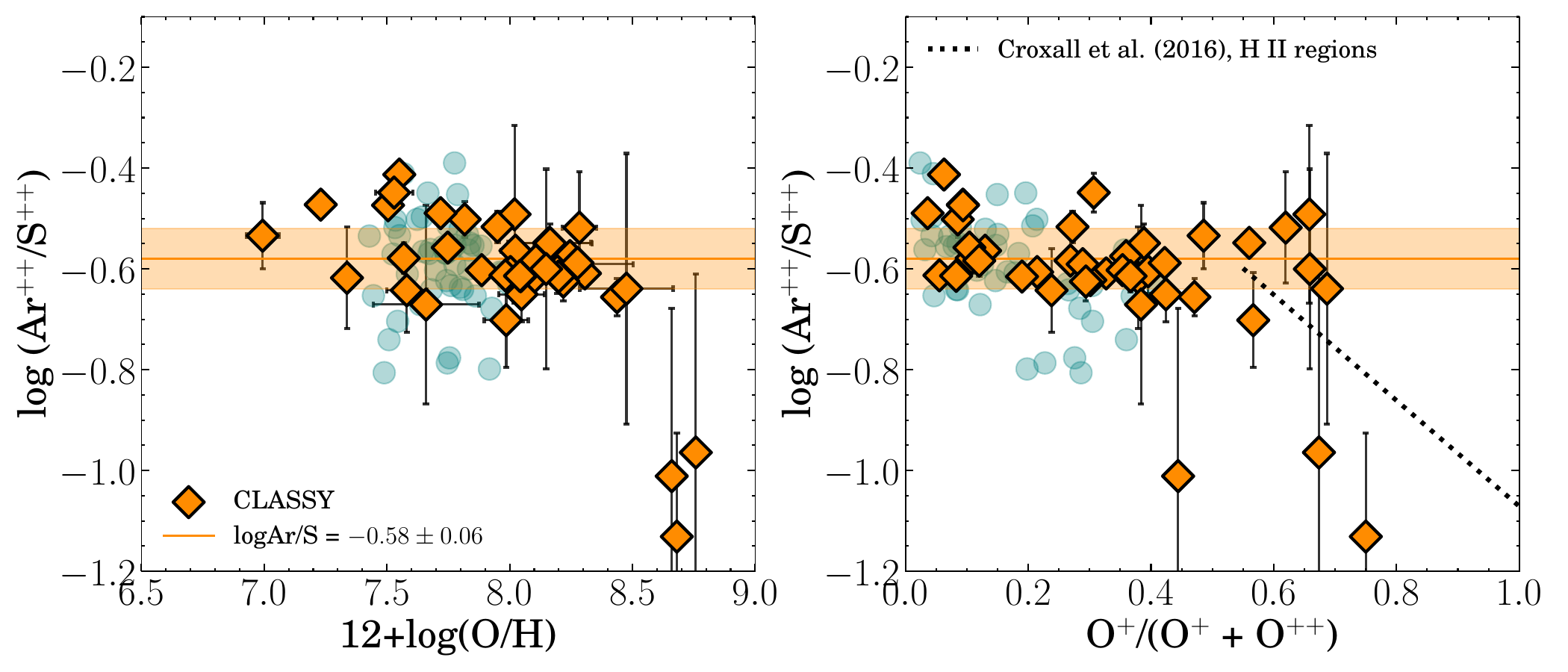}
        \caption{The ionic abundance ratio of Ar$^{2+}$/S$^{2+}$ as a function of metallicity (\textit{left}) and O$^{+}$/O (\textit{right}) for CLASSY. The circles are SFGs of the sample described in Section~\ref{sec:sample}. The mean and standard deviation value of log(Ar$^{2+}$/S$^{2+}$) is illustrated with a solid line color-coded with CLASSY. The result of log(Ar$^{2+}$/S$^{2+}$) $=-0.58\pm0.06$ is in agreement with those found in \hii\ regions \citep{kennicutt03a, croxall16} at O$^{+}$/O $< 0.60$. While the value of log(Ar$^{2+}$/S$^{2+}$) for CLASSY stays constant, low ionization \hii\ regions show a significant decrement at high values of O$^{+}$/(O$^{+}$ +O$^{2+}$) $\ge 0.60$ \citep{croxall16}. This decrement is related to the increased contribution of Ar$^{+}$ in low ionization objects. The green solid line quantifies the decrement derived by \citet{croxall16} for \hii\ regions in M101.}
\label{fig:S_Ar_ionic}
\end{center}
\end{figure*}

\subsubsection{Ar$^{2+}$/S$^{2+}$ $\sim$ Ar/S}
%
We investigated the behavior of the abundance ratio of Ar$^{2+}$/S$^{2+}$. 
With similar ionization potentials of 27.6 eV and 23.3 eV for Ar$^{2+}$ and S$^{2+}$, 
respectively, these two ions both trace the intermediate ionization state of the gas 
\citep[e.g.,][]{berg21}. 
In Figure~\ref{fig:S_Ar_ionic}, we present log(Ar$^{2+}$/S$^{+2}$) with respect to metallicity
(left panel) and ionization degree (right panel) for both the CLASSY sample
and the comparison sample. 
For the CLASSY sample, we calculate a mean of log(Ar$^{2+}$/S$^{2+}$) $= -0.58\pm0.06$,
which is in excellent agreement with the values reported in \citet{kennicutt03a} and 
\citet{croxall16} for individual \hii\ regions in M101. 
However, \citet{croxall16} showed that Ar$^{2+}$/S$^{2+}$ decreases significantly in 
low ionization objects (O$^{+}$/(O$^{+}$+ O$^{2+}$) $\ge$ 0.6, see also Figure~\ref{fig:S_Ar_ionic}) 
due to the contribution of Ar$^{+}$ (dotted line in Figure~\ref{fig:S_Ar_ionic}). 
The CLASSY galaxies stay constant at larger values of O$^{+}$/O for most of the sample. 
This implies that the Ar$^{2+}$/S$^{2+}$ abundance ratio serves as a good approximation for 
the total Ar/S ratio, Ar$^{2+}$/S$^{2+}$ $\sim$ Ar/S (i.e., removing the use of the ICFs 
due to the similar ionization potentials).

While the Ar$^{2+}$/S$^{2+}\sim$Ar/S trend is relatively constant and the Ar/O trends from
[\ion{Ar}{3}]$+$[\ion{Ar}{4}] are similarly well behaved in Figure~\ref{fig:alpha_elements},
the S/O trends are more messy due to large ICF uncertainties 
(see discussion in Section~\ref{sec:sulfur}).
Therefore, our analysis of Ar/S suggests that the most probable explanation for the low 
values of S/O, and subsequent large dispersion, is due to the ICF of S used. 
This implies that the adopted \citet{izotov06} ICF might be useful for individual \hii\ region
abundances, but may not be appropriate for integrated galaxy spectra. 
\color{black}

\begin{deluxetable}{l c c  c  C} 
\tablecaption{Mean and standard deviation values of Ne/O, S/O, Cl/O, and Ar/O for the CLASSY sample galaxies}
\label{tab:mean_dispersion} 
\tablewidth{0pt}
\tablehead{
log$X$/O   &   $N$    &  $\mu\pm\sigma$  & Lines & ICF}
\startdata
Ne/O    &  34  & $-0.63\pm0.06$ & [\ion{Ne}{3}] & 1 \\
S/O    &  31  & $-1.73\pm0.10$   & [\ion{S}{2}] + [\ion{S}{3}] & 2\\
Cl/O    &  15  & $-3.60\pm0.17$ & [\ion{Cl}{3}] & 2\\
Ar/O    &  20  & $-2.46\pm0.18$ & [\ion{Ar}{3}] & 2\\
Ar/O    &  22  & $-2.36\pm0.09$ & [\ion{Ar}{3}] + [\ion{Ar}{4}] & 2 \\
\enddata
\tablecomments{ (1) \cite{dors13}, (2) \citet{izotov06}. For the S/O ratios, we discard those galaxies with high S/O. The mean and dispersion values considering only the measurement of [\ion{Ar}{3}]  for 42 galaxies is $-2.43\pm$0.16.}
\end{deluxetable}

\subsection{Total vs. Relative Abundances}
\label{sec:relative-abundances}
\begin{figure*}
\begin{center}
    \includegraphics[width=0.95\textwidth, trim=0 0 0 0,  clip=yes]{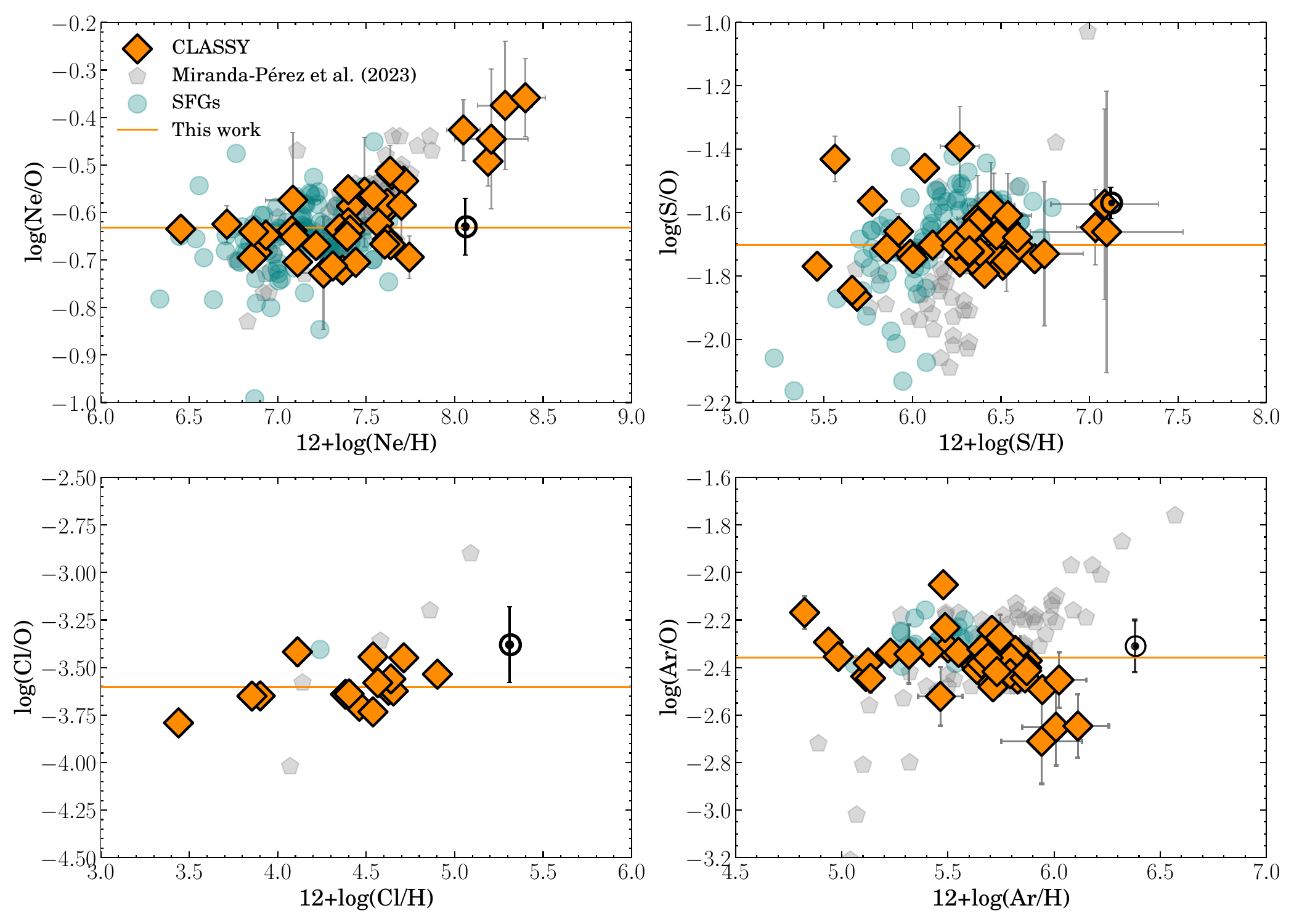} \caption{The relative abundances of Ne, S, Cl, and Ar with respect to Ne/O, S/O, Cl/O, and Ar/O.  The circles and pentagons represent the sample of SFGs compiled in this study and  85 $z \sim0$ SFGs of \citet{mirandaperez23}. The solar symbol in each plot indicates the photosphere abundance of Ne, S, Cl, and Ar of \citet{asplund21}. The solid line indicates the fit of the abundance ratios for CLASSY (see Figure~\ref{fig:alpha_elements}). The abundances of the sample of SFGs of \citet{mirandaperez23} were determined using different ICFs than those implied in this study. The Ne/O and Ar/O ratios show a significant trend with respect to the relative abundances of Ne and Ar. }
\label{fig:comparsion-sfgs}
\end{center}
\end{figure*}

In this section, we present the total abundances of Ne, S, Cl, and Ar 
and compare to their respective relative $\alpha$-abundances. 
In Figure~\ref{fig:comparsion-sfgs}, we show the comparison of the Ne/O abundance ratio derived for the CLASSY galaxies. In addition, we have added the results of \citet{mirandaperez23}, who derived multiple chemical abundances in a sample of SFGs from the SDSS. For Ne/O, \citet{mirandaperez23} used the ICF of \citet{torres-peimbert89} (Ne$^{2+}$/O$^{2+}$ $\approx$ Ne/O), which is based on the similarity of the ionization potentials of Ne$^{2+}$ and O$^{2+}$.

The results of \citet{mirandaperez23} also show that the Ne/O ratio increases with metallicity at 12+log(O/H) $\sim$ 8.21 \citep[see also][]{izotov11}. In fact, for CLASSY and the sample from the literature (SFGs and \ion{H}{2} regions), we obtained similar results showing a constant trend with metallicity (see Figure~\ref{fig:alpha_elements}). 
Although, it is unclear what is the role of the ICF(Ne) at a low ionization degree, the ICFs in SFGs of Ne should be assessed, in particular, in metal-rich galaxies (12+log(O/H) > 8.2). However,  chemical evolution models predict a secondary production of Ne at high metallicities in a similar way to N. In such as scenario Ne increases more quickly than O above the solar value \citep[see Figure 39 of][]{kobayashi20}.

We also compare S/O with respect to the sample of \citet{mirandaperez23}. Again, in Figure~\ref{fig:comparsion-sfgs}, we show the S/O \textit{vs.}\ S/H for the CLASSY galaxies. In fact, for the set of galaxies of  \citet{mirandaperez23} the results show lower values of S/O in comparison with CLASSY (12+log(S/H) = 5.5-6.5), while a few objects show high values of S/O. \citet{mirandaperez23} also report a strong trend of S/O with metallicity, but it is unclear how many galaxies with measurements of [\ion{S}{2}] and [\ion{S}{3}] are used to derive S/O and S/H, since the use of only S$^{+}$ ([\ion{S}{2}] \W\W6717,31) can introduce a strong dependence of S/O with metallicity \citep[see][]{amayo21}. However, we note that the sample of SDSS galaxies shows lower abundances relative to the solar value, in agreement with the S/O trends of the CLASSY galaxies (see also Figure~\ref{fig:alpha_elements}). 

In the bottom panel of Figure~\ref{fig:comparsion-sfgs}, we also compare the Cl/O ratios of the CLASSY galaxies with the results of \citet{mirandaperez23}. While with CLASSY we get a constant trend with Cl/H, the sample of \citet{mirandaperez23} shows a slight trend of Cl/O as Cl/H increases. However, the dispersion in the Cl/O \textit{vs.}\ O/H relationship (see also Figure~\ref{fig:alpha_elements}) plus the few measurements in the sample \citet{mirandaperez23} provide a less robust comparison.

As an additional comparison, in Figure~\ref{fig:comparsion-sfgs} we show the Ar/O ratios with respect to Ar/H for CLASSY and the results of \citet{mirandaperez23}. We find a good agreement with the sample of \citet{mirandaperez23} around the range of 12+log(Ar/H) = 5.3-5.9. However, at high values of Ar/H, we found an opposite behavior than \citet{mirandaperez23}, whose sample shows a significant trend with the Ar/H abundance. An opposite behavior is also found at lower values of 12+log(Ar/H) $\sim$ 5. 
Recently, \citet{arnaboldi22} analyzed the abundance pattern of the O/Ar \textit{vs.}\ Ar/H relation for a sample of planetary nebulae in M31. \citet{arnaboldi22} found that the highest values of O/Ar ($\sim$2.5) are the lowest values of Ar/H $\sim 6$. With CLASSY, there are only a few galaxies with Ar/H $\ge 6$, and such a correlation is unclear. However, the sample of galaxies of \citet{mirandaperez23} shows that Ar/O increases as the total abundance of Ar increases. A significant correlation between Ar/O and Ar/S is not expected due to origin of these elements. However, chemical evolution models predict that such a correlation might be due to the additional contribution of Ar from Type-Ia SNe \citep{kobayashi20, kobayashi20b, arnaboldi22}. Such a scenario suggests that the mechanism production of Ar can be due to both CCSNe and Type-Ia SNe, although a large time delay is expected until Ar be enrichment for this mechanism.    

At lower values of Ar/H (and low metallicity) the trend with Ar/O is constant as expected for the nucleosynthesis origin of Ar and O, which are produced by CCSNe (see Figs~\ref{fig:alpha_elements} and \ref{fig:comparsion-sfgs}). For the lowest values of Ar/H $\sim$ 5, the two different trends shown in CLASSY and the sample of \citet{mirandaperez23} might be related to the ICF and its performance in the integrated spectra of galaxies with a lower degree of ionization. 


In general, the abundance patterns of Ne, S, and Ar of the CLASSY galaxies follow the expected trends for nucleosynthesis (with important bias at high metallicities). Another possible cause of the behavior is that the abundances of Ne, S, and Ar might be affected by the interplay between stellar yields and the stellar life of massive stars \citep{matteucci05}. Using chemical evolution models \citet{matteucci05} showed that the abundance ratio of [O/S] and [O/Si] (relative to the solar abundance) vary as a function of metallicity due to the non-negligible contribution of Type Ia SNe \citep{matteucci05, arnaboldi22, kobayashi20}. Therefore, while oxygen is uniquely produced by massive stars, the production of $\alpha-$elements is not constant due to the whole range of masses that those elements are produced \citep{matteucci05} as the metallicity increases. Chemical evolution models in the solar neighborhood also predict the expected constant behavior for S (and similar for Ne and S) but with a slight decrease of this abundance as metallicity increases (i.e., [Fe/H] in stars) \citet{kobayashi20}.

\section{Abundance patterns in the early Universe}
\label{sec:high-z}
With \textit{JWST} providing deep spectra of high-redshift galaxies, 
the first Ne, S, and Ar abundances have been provided for the early universe
\citep[e.g.,][]{arellanocordova22a, isobe23a, marqueschaves23, Rogers24}. 
Using these studies, we can investigate the evolution of relative $\alpha$-abundance 
patterns of O, Ne, S, and Ar for the first time.
In particular, \citet{isobe23a} derived Ne/O, S/O, and Ar/O abundances for a sample 
of 70 $4<z<10$ SFGs compiled from the ERO \citep{pontoppidan:22}, GLASS 
\citep{treu22}, and CEERS \citep{finkelstein22a} programs.
\citet{isobe23a} used direct \Te\ measurements for 13 galaxies, but had to assume a 
\Te\ = 15000$\pm5000$ K and $n_{\rm e} = 300$ cm$^{-3}$ for the remaining 57 galaxies. 
We, therefore, augment the \citet{isobe23a} sample with direct \Te\ measurements for 
a $z\sim11$ galaxy from \citet{marqueschaves23},
three $z>7$ galaxies from \citet[][replacing three galaxies reported in Isobe et al.]{arellanocordova22a}, 
and a $z>3$ galaxy from the CECILIA survey \citep{strom23,Rogers24}.
In order to consistently compare the CLASSY abundance patterns with $z > 3$ galaxies, 
we require direct measurements of \Te, derived using either [\ion{O}{3}]~\W4363 or 
\ion{O}{3}]~\W1666, and \Ne, derived from [\ion{O}{2}] \W3729/\W3726, with S/N $\geq$ 4.

\subsection{Cosmic evolution of Ne/O}
\begin{figure*}
\begin{center}
    \includegraphics[width=0.95\textwidth]{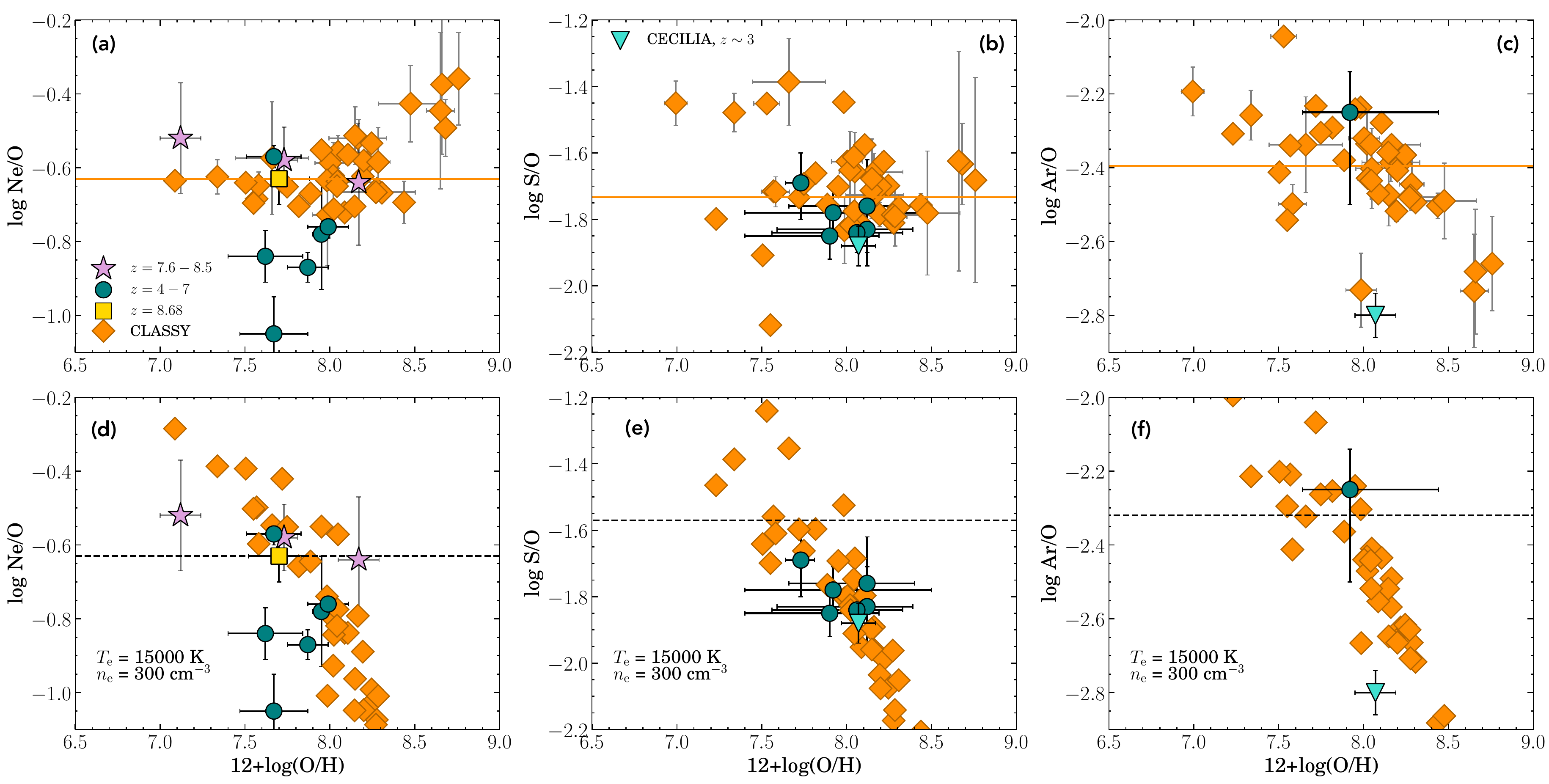}
        \caption{
        {\it Top:} Relative $\alpha$-abundances versus oxygen abundance for the 
        CLASSY sample ($z\sim0$; orange diamonds) in comparison with SFGs at $z>3$ from 
        \citet[][turquoise triangle]{Rogers24}, \citet[][pink stars]{arellanocordova22a}, 
        \citet[][green circles]{isobe23a}, and \citet[][gold squares]{marqueschaves23}.
        The error-weighted mean $\alpha$-ratios are indicated by the solid orange lines in each panel
        (see also Section~\ref{fig:alpha_elements}).
        Panel (a) shows the Ne/O trend, panel (b) shows the S/O trend, and panel (c) shows 
        the Ar/O trend, with an additional $z\sim0$ comparison sample from 
        \citet[][blue pentagons]{Kojima2021}.  
        Note that the abundance patterns of S/O and Ar/O of \citet{isobe23a} were estimated using \Te\ = 15000$\pm$5000 K and $n_{\rm e} = 300$ cm$^{-3}$. 
        {\it Bottom:} The same abundance patterns as shown in the top row, but with CLASSY abundances 
        recalculated assuming \Te\ = 15000$\pm$5000 K and $n_{\rm e} = 300$ cm$^{-3}$ to match the 
        method of the high-$z$ sample in \citet{isobe23a}. 
        While the low- and high-$z$ trends overlap, the bottom row plots show significant decreasing
        trends with increasing metallicity and larger dispersions than the trends in the top row. 
        This result shows the importance of characterizing the temperature structure of the gas in SFGs.      }
\label{fig:high-z-comp}
\end{center}
\end{figure*}

In Figure~\ref{fig:high-z-comp} (a), we compare the Ne/O abundance pattern of CLASSY 
with the 10 $z>3$ SFGs that meet of direct-\Te\ requirement and have Ne/O measurements.
We found that at $z>3$ the average log(Ne/O) = $-$0.72$\pm$0.17.
In comparison to the average log(Ne/O) $=-0.63\pm0.06$ for the CLASSY sample, the higher redshift
galaxies show similarly large scatter but sub-solar Ne/O abundances.
One the other hand, the $z = 8.678$ SFG (yellow square) shows an excellent agreement with 
the solar abundance and the mean values of Ne/O for CLASSY. 
This result is consistent with the lack of evidence of cosmic evolution of Ne/O as reported in 
\citet{arellanocordova22a} for z $> 7$ galaxies using JWST/NIRSpec ERO observations 
\citep{pontoppidan:22, trump23}. 
The low log(Ne/O) $< -1.0$ values in the \citet{isobe23a} sample are explained in that work using  
chemical evolution models of \citet{watanabe23}. These authors concluded that the models with $ M \ge $ 30 $M_{\odot}$ can reproduce the low values of Ne/O since Ne is reduced due to the high temperature in the carbon-burning layer of these stars, which should be typical at high-$z$ \citep[see,][]{isobe23a}. 

Low values of Ne/O are also observed in local SFGs, but typically only from \ion{H}{2} region 
spectra at high-metallicities (12+log(O/H) $>8.0$; see Figure~\ref{fig:alpha_elements}). 
The cause of these low values of Ne/O is still unclear but it might be related to the ionization 
structure of the gas, or to ICFs that have been poorly calibrated for the extreme conditions and 
integrated spectra of high-$z$ galaxies. 
Therefore, to evaluate this discrepancy in the abundance patterns of Ne/O, it is crucial to 
increase the number of galaxies at $z>3$ with accurate \Te-based abundances. 
In this context, the present study with CLASSY provides a robust foundation to compare the Ne/O 
abundances of high-$z$ galaxies and evaluate its chemical evolution across cosmic time.

\subsection{Cosmic Evolution of S/O}
Figure~\ref{fig:high-z-comp} (b) shows the evolution of the S/O abundance pattern.
Unfortunately, only two high-$z$ galaxies have \Te-based S/O measurements 
(1 from \cite{isobe23a} and 1 from \cite{Rogers24}).
Thus, we also consider galaxies from \citet{isobe23a} with S/O measurements but no \Te\ detection 
in order to increase our sample of high-$z$ galaxies to seven. 
Note that we have added only those galaxies whose uncertainties are also reported 
(i.e., avoiding upper limit results). 
In this context, it is crucial to be careful with the interpretation of relative abundances in 
our comparison of CLASSY with high-$z$ galaxies due to the missing detection of \Te. 
However, these results can provide a general view of the high-$z$ abundance distributions. 

Interestingly, the high-$z$ sample in Figure~\ref{fig:high-z-comp} (b) shows very little scatter
and is in good agreement with the average $z\sim0$ CLASSY trend. 
In particular, the mean S/O abundance of the high-$z$ sample (log(S/O) $= -1.88\pm0.06$) is 
consistent with the average value of the CLASSY sample (log(S/O) $= -1.73\pm0.10$), suggesting there is no cosmic evolution
of the S/O abundance. 
However, it is important to mention that the high-$z$ S/O values are computed using only the 
measurements of [\ion{S}{2}] $\lambda\lambda6717,31$ \citep[see Figure B1 of ][]{isobe23a}. 
The use of solely the low-ionization S lines might introduce significant correlations in the 
S/O trend with the ionization parameter and metallicity \citep[e.g.,][]{amayo21}. 

\subsection{Cosmic Evolution of Ar/O}
Here we consider the cosmic evolution of the Ar/O--O/H trend.
Similar to S/O, only the $z\sim3$ galaxy from \citet{Rogers24} has a \Te-based Ar/O abundance.
We again consider galaxies from \citet{isobe23a} with Ar/O measurements but no 
\Te detection, however, this only adds one galaxy to the high-$z$ sample. 
With no significant sample of high-$z$ Ar/O abundances, we limit our analysis of 
Ar/O to the abundance from \citet{Rogers24} but still show a visual examination in
Figure~\ref{fig:high-z-comp} (c).

Interestingly, the Ar/O ratio reported in \citet{Rogers24} at $z\sim3$ is significantly lower 
than the average $z\sim$0 value of the CLASSY sample. 
However, Figure~\ref{fig:alpha_elements} also shows that four CLASSY galaxies 
(J0808+3948, J0940+2935, J1144+4015, J1525+0757)
have similar values of Ar/O as in \citet{Rogers24} for a similar 12+log(O/H) $> 8.0$. 
A detailed inspection with a large sample of high-$z$ galaxies (at moderate metallicities) 
is needed to evaluate the Ar/O evolution.
The Ar/O ICFs may also require updated calibrations for integrated galaxy spectra.

\subsection{Temperature effects in abundance determinations at high redshift}
\label{appen:text_te_OH}

The increase of $z>3$ galaxies with the emission lines necessary to derive 
$\alpha$-element abundances 
is an extraordinary opportunity to investigate the relative chemical enrichment of Ne, S, and Ar. 
However, some high-$z$ systems lack the auroral line detections necessary for a 
robust \Te\ measurement. 
Accurate chemical abundance determinations requires \Te\ measurements
since collisionally excited lines (e.g., [\ion{Ne}{3}], [\ion{Ar}{3}], and [\ion{S}{3}]) 
depend exponentially on \Te\ \citep[e.g.,][]{osterbrock06, peimbert17}. 

Most high-$z$ abundance measurements still lack \Te\ measurements
and must, therefore, assume fixed \Te\ and \Ne\ values. 
In this context, it is important to assess the bias introduced relative and total abundance
determinations by unknown temperatures.
To evaluate this bias, we re-calculated the abundance patterns of the CLASSY galaxies following the procedure of \citet{isobe23a} when \Te\ is not available. 
This procedure uses fixed values of \Te = 15000 K and \Ne = 300 cm$^{-3}$.  

In the bottom row of Figure~\ref{fig:high-z-comp}, we show the Ne/O, S/O, and Ar/O trends 
as a function of O/H for the CLASSY sample assuming fixed \Te\ and \Ne\ values. 
Compared to the \Te-abundances in the top row, the fixed-\Te\ values show much strong
correlations with metallicity.
These trends also extend to lower Ne/O, S/O, and Ar/O values, suggesting that fixed
temperatures may underestimate relative $\alpha$-abundances.
Therefore, the implications of comparing samples with abundances derived using different 
\Te\ methods/assumptions can lead to significant bias in our interpretation of the chemical 
enrichment history. 
Therefore, we stress that in order to accurately evaluate the chemical evolution of galaxies, 
a sample of galaxies with high-S/N detections and robust measurements of the \Te-sensitive 
[\ion{O}{3}], [\ion{S}{3}], and/or [\ion{N}{2}] lines are necessary to reduce possible bias. 

\section{Scaling relations of metals}
\label{sec:scaling-relations}
Here, we discuss the abundance patterns of O, Ne, S, Cl, and Ar as a function of the galaxy 
properties such as stellar mass and SFR for CLASSY in comparison with the results of 
$z > 3$ galaxies. 

\subsection{Mass-Metallicity Relation (MZR)}
The mass-metallicity relation (MZR) is shaped by evolutionary processes associated with the 
baryon cycle and so provides essential information on the growth and 
evolution of galaxies \citep[e.g.,][]{lequeux79, tremonti04, maiolino+19, curti23b, nakajima23}. 
Here, we compare the MZR of CLASSY with our high-$z$ ($z>3$) galaxy sample. The SFR and stellar masses were compiled from the original papers. 
In Figure~\ref{fig:mzr}, we show the mass-metallicity relation (MZR) for CLASSY derived in \PI). 

First, we compare these results with other local SFGs. 
In the top panel of Figure~\ref{fig:mzr}, we plot the slope (magenta dashed line) derived 
using the Low-Mass Local Volume Legacy (LMLVL, $z \sim 0$) of \citet{berg12} with \Te\ direct 
abundances. 
The CLASSY sample follows the trend derived for LMLVL SFGs, with a similar slope and 
with a slight offset to higher metallicities (see also \PI). 
In Figure~\ref{fig:mzr}, we have also added as pink stars the results of $z > 7$ galaxies analyzed in \citet{arellanocordova22b}. 
As shown in \citet{arellanocordova22b}, there is no evidence for evolution of the MZR given 
that the $z > 7$ galaxies follow the local MZR with the dispersion ($\sigma = 0.29$).

Recently, some authors have focused on understanding the physical processes driving the MZR, and its evolution across cosmic time using a statistical sample  of $z > 3$ galaxies \citep[e.g.,][]{curti23b, nakajima23}. 
Here, we compare the slopes derived in \citet{nakajima23} and \citet{curti23b}, who used samples of SFGs obtained from CEERS \citep{finkelstein22a}, GLASS \citep{treu22}, and JADES \citep{eisenstein23}, respectively. 
The range of valid masses for these slopes is 
$M_{*}$ < 10$^{9.5}$ M$_{\odot}$. It is important to note that the slopes derived in such studies depend on the empirical calibrators \citep[with few objects with \Te-metallicities, e.g.,][]{nakajima23}. Both samples cover similar ranges in redshift ($z=3-10$). 
The $z = 4-10$ MZR derived by \citet[][dashed green line]{nakajima23} follows the trend of CLASSY galaxies and the LMLVL sample with a slight offset to lower metallicities.
On the other hand, the MZR of galaxies at $z>6$ is flatter than that of the $z\sim0$  
CLASSY and LVLML samples.
Similarly, \citet{curti23b} found a relatively flat MZR slope for their full $z=3-10$ sample 
(0.17$\pm$0.03), and a significantly flatter slope (0.11$\pm$0.05) for the highest redshift galaxies in their sample ($z=6-10$).
 Such flat slopes have been attributed to a different feedback mechanism dominating in dwarf 
M$_{\star} < 10^{9.5}$ M$_\odot$ $z \sim 2-3$ galaxies \citep{li23}.

Since a steep slope is seen for all masses in the the local CLASSY \Te-method MZR, 
the shallower slopes of the $z>3$ MZRs may be the result of using empirical calibrations.
On the other hand, the $z>3$ MZR slopes are compatible with the results derived for $z\sim0$ 
green pea and blueberry galaxies using the direct method \citep{yang17}, which systematically 
differs from our CLASSY results. 
Therefore, a significant sample of $z \sim 3$ galaxies with direct metallicities are fundamental to constrain the MZR to the mild/strong evolution of the MZR at those redshifts \citep[e.g.,][]{sanders21, strom23, Rogers24}. 
Such a sample could then improve the performance of empirical calibrations, which will play a crucial role in understanding the evolution and shape of the MZR across cosmic time\citep[e.g.,][]{kewley08, patricio16}.

\subsection{Fundamental Metallicity Relation (FMR)}
The relation between stellar mass, metallicity, and SFR is the so-called fundamental metallicity relation \citep[][FMR]{mannucci10}, due to the secondary dependence of the MZR \citep{ellison08}. The FMR was first characterized for $z =0-4$, and it is associated with important evolutionary processes in galaxy formation, primarily the infall of pristine gas which acts to elevate the SFR at the same time as diluting the gas-phase metallicity
\citep[e.g.,][]{mannucci10, laralopez10, andrews13, sanders21, kumari21b}. 

\begin{figure*}
\begin{center}
    \includegraphics[width=0.98\textwidth]{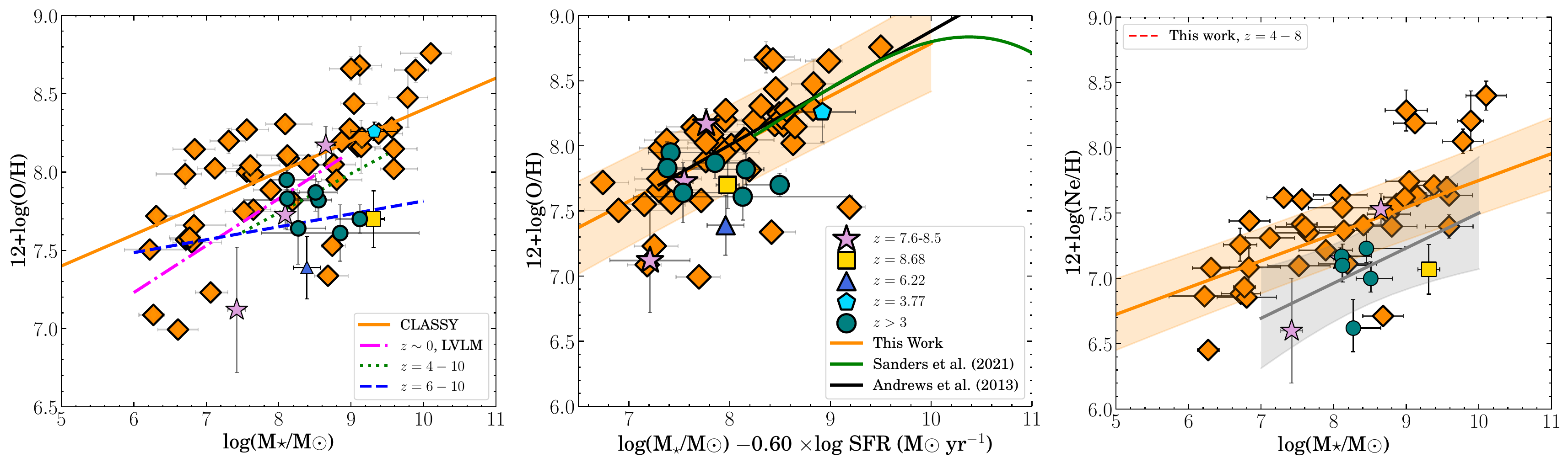}
        \caption{
        {\it Left:} The evolution of the MZR.
        The $z\sim0$ CLASSY sample is shown as orange diamonds, with the best-fit MZR 
        from \citet{Berg22} shown as an orange line. 
        In comparison, we also plot high-$z$ galaxies from
        \citet[$z>7$; ][pink stars]{arellanocordova22b}, 
        \citet[$z>4$;][teal circles]{nakajima23}, and
        \citet[$z\sim9$][yellow square]{marqueschaves23} and
        MZRs from \citet[][$z=0$]{berg12}, \citet[][$z = 4-10$]{nakajima23}, and 
        \citet[][$z=6-10$]{curti23b}.
        {\it Center:} The evolution of the FMR for CLASSY in comparison to  
        $z >$ 3 galaxies with \Te-direct abundance determinations. 
        The green line is the projection fit of the FMR of \citet[at $z = 0-3$]{sanders21} and 
        \citet[at $z \sim0$]{andrews13} using stacked spectra with metallicity determined using strong-line methods and the 
        \Te-sensitive method, respectively. 
        CLASSY and $z>3$ galaxies follow the FMR relation with no apparent evolution.
        {\it Right:} Evolution of the MNeR. 
        The solid line represents the best fit to the CLASSY data, while the dashed line 
        illustrates the fit to the high-$z$ galaxy sample. 
        There appears to be an evolution of the MNeR:
        higher redshift galaxies have both a steeper MNeR slope of 0.27$\pm$0.19 and 
        a lower y-intercept.  \label{fig:mzr}}
\end{center}
\end{figure*}

The functional form of the FMR was defined by \citet{mannucci10} as 
\begin{equation}
\label{eq:mu_fmr}
    \mu_{\alpha} = \log(M_\star) - \alpha\cdot\log({\rm SFR}), 
\end{equation}
where stellar mass and SFR have units of $M_\odot$ and $M_\odot$ yr$^{-1}$, respectively. 
Adopting this equation, we determine the best fit FMR to the CLASSY data to have
$\alpha=60$ such that:
\begin{equation}
    12+\log(\rm O/H) = (0.40\pm0.08)\times\mu_{60} + (4.7\pm0.65), 
\end{equation}
with a standard deviation of $\sigma = 0.32$ dex.
We illustrate our derived FMR in in the middle panel of Figure~\ref{fig:mzr} in 
comparison to $z > 3$ galaxies with metallicities determined using the direct method. 
The results of \citet{arellano-cordova2020b}, \citet{jones23}, \citet{marqueschaves23}, and 
\citet{nakajima23} illustrate observations of \textit{JWST} for $z > 3$ galaxies, 
while the result of \citet{citro23} corresponds to ground-based observations of 
a lensed galaxy at $z\sim4$. 

We also compare to the FMR relations of \citet{sanders21} and \citet{andrews13}. 
Note, however, both \citet{sanders21} and \citet{andrews13} derived their FMRs using stacked 
spectra of $z \sim 0$ galaxies, but find $\mu_{\alpha} = 0.60$ and 0.66, respectively,
in good agreement with the FMR for CLASSY galaxies. 
Interestingly, the CLASSY galaxies with the lowest metallicities (12$+$log(O/H)$\sim7.5$) depart 
from the general trend of the sample and the FMR, 
(J0405$-$3646, J1132$-$5722, J0127$-$0519, J0337$-$0502, and J934+5514). 
although three of these galaxies (J0337$-$0502, J0405$-$3646 and J934+5514) have properties 
that match well with the properties of $z>7$ galaxies in \citet{arellanocordova22b} within 
the uncertainties. 

In general, the $z > 3$ sample shows good consistency with the $z=0$ CLASSY FMR, suggesting
not evolution in the FMR for $z=0-3$,
Recently, \citet{nakajima23} analyzed the cosmic evolution of the FMR relation for $z = 4-10$ galaxies. These authors reported that galaxies at $z < 8$ are consistent with the local FMR but find evidence for an evolution at $z > 8$, suggesting a fundamental change in the physical processes acting on galaxies at the earliest epochs.
Similarly, \citet{curti23b} report an evolution away from the local FMR for galaxies at $z > 6$.

Although we do not see strong evidence for an FMR evolution in the middle plot of Figure~\ref{fig:mzr}, the sample size is small. 
As before, however, it is important to note that the results of \citet{curti23b} and \citet{nakajima23} are based on empirical metallicity calibrations that can be strongly affected by uncertainties in the ionization and temperature structure of the \hii\ regions at high redshift. Larger samples of \Te-based metallicities at $z > 3$ will again be crucial in robustly characterising the FMR evolution at the highest redshifts \citep[e.g.,][]{stasinska10, patricio16, nakajima22, nakajima23}. 

\subsection{The mass-neon relation (MNeR)}

Since Ne has the same nucleosynthesic origins as O, 
the total Ne/H abundance also correlates with the stellar mass. 
In the bottom panel of Figure~\ref{fig:mzr}, we present the mass-neon 
relation (MNeR) for CLASSY. 
The solid line in Figure~\ref{fig:mzr} represents the unweighted linear least squares fit to the CLASSY data, and it is presented in the following expression:
\begin{equation}
    {\rm 12+log(Ne/H)} =  (5.93\pm0.33) + (0.18\pm0.04) \times \log M_\star, 
\end{equation}
with a standard deviation of $\sigma=0.22$ and log $M_\star$ in units of $M_\odot$.
Note that we have omitted metal-rich galaxies with high values of Ne/O in CLASSY (see Figure~\ref{fig:alpha_elements}). 
We have added the sample of $z > 4$ galaxies from \citet{arellanocordova22a}, \citet{isobe23a}, and \citet{marqueschaves23} with direct \Te\ abundance determinations.

We measure the first MNeR at high redshift, finding an unweighted least-squares 
linear trend of:
\begin{equation}
    {\rm 12+log(Ne/H)} =  (4.81\pm1.58) + (0.27\pm0.19) \times  \log M_\star, 
\end{equation}
with a standard deviation of $\sigma=0.24$ and log $M_\star$ in units of $M_\odot$.
We also provide in Table~\ref{tab:relationships-properties} the best fits to S, Cl, and Ar with respect to the stellar mass, showing similar slopes of $\sim$ 0.20-0.24 (see Figure~\ref{fig:total-properties} in Appendix~\ref{apped:galaxy-properties-elementes}.) 

The MNeR in figure~\ref{fig:mzr} suggests some evolution, but the lower Ne/H tend of the $z >4$ galaxies are consistent within the MNeR uncertainties. 
The differences between $z > 7$ galaxies of \citet[][purple stars]{arellanocordova22a} and $z=4-8$ galaxies of \citet[][teal cirlces]{isobe23a} might be due to the ICFs or the presence of a young population of massive stars \citep{isobe23a}. 
On the other hand, CEERS-1019 of \citet{marqueschaves23} is in agreement with the sample of galaxies of \citet{isobe23a}, showing a lower Ne/H abundance than the local SFGs. 
The dispersion in the high-$z$ MNeR is comparable to what we find in CLASSY (see Table~\ref{tab:mass-abundances-relations}). 
Although, this is the first assessment of the Ne-MR at high$-z$, robust measurements of Ne/H and Ne/O with \Te\ and a robust ICF are necessary to better interpret the shape of the MNeR and confirm or refute the lower values of Ne/O at $z > 6$.

\begin{deluxetable}{l c c  c c } 
\tablecaption{The mass and abundance pattern relations of O, Ne, S, Cl, and Ar for CLASSY galaxies}
\label{tab:mass-abundances-relations} 
\tablewidth{0pt}
\tablehead{Ion   &  $\alpha$   &  12+log($X$/H)$_{0}$ & $\sigma$}
\startdata
Ne    &  0.20$\pm$0.04 &  5.93$\pm$0.31 & 0.22 \\
S     &  0.24$\pm$0.04 &  4.32$\pm$0.33 & 0.28 \\ 
Cl    &  0.24$\pm$0.09 &  2.47$\pm$0.75 & 0.34 \\ 
Ar    &  0.20$\pm$0.03 &  3.97$\pm$0.23 & 0.23 \\ 
\enddata
\tablecomments{
$\alpha$ represents the slope and 12+log($X$/H) is the total abundance of each element labeled in column 1. 12+log($X$/H) = $\alpha$M$_{*}$ + 12+log($X$/H)$_{0}$. $\sigma$ represents the dispersion of the data with respect to the linear fit. } 
\label{tab:relationships-properties}
\end{deluxetable}

\section{Summary and Conclusions} 
\label{sec:conclusion}
We investigated the abundance patterns of Ne, S, Cl, and Ar using a sample of 43 SFGs from the CLASSY survey with significant \Te\ measurements. 
The CLASSY sample has enhanced SFRs compared to other local sample of SFGs,
making it a useful reference sample for high-redshift galaxies.
We used different sets of \Te-diagnostics to provide a robust characterization of the different ionization and temperature structures of the gas for each galaxy. %
Further, we analyzed a set of literature ICFs and determined the most appropriate ICFs for use with the CLASSY galaxies. 
The detailed inspection of the ICFs, \Te, and ionization structure of the CLASSY galaxies reduces biases related to the scatter and systematic trends with metallicity. 
As a result, we present robust measurements of abundance patterns of Ne/O, S/O, Cl/O, and Ar/O, and the total abundances of Ne, S, Cl, and Ar for CLASSY. 

We examine the abundance patterns of CLASSY compared to two samples of local star-forming systems: (i) individual \ion{H}{2} regions and (ii) integrated galaxies, to access the appropriateness of local calibrations for higher-redshift samples.
We then compared to a high-redshift literature sample ($z > 3$) of abundances to investigate the chemical evolution of Ne, S, Cl, and Ar. Finally, the broad galaxy properties covered in CLASSY (see \PI\ and \PIV) allow us to analyze essential scaling relations related to galaxy evolution.
Our main conclusions can be summarized as follows:

\begin{itemize}

    \item With the accurate determination of the ionic abundances of the CLASSY galaxies, we carefully inspect the ICFs for Ne, S, Cl, and Ar. We find that the ICF for Ne of \citet{dors13} provides less scatter and an excellent agreement with the solar abundance. For the rest of the elements (S, Cl, and Ar), the ICFs of \citet{izotov06} show less dispersion in comparison with other sets of ICFs examined in this study \citep[e.g.,][]{amayo21, thuan95, dors16}. However, we emphasize that the application of these ICFs needs to be done with care since these ICFs are not representative of the integrated spectra of SFGs. However, the size of the ICF scatter is not enough to explain the unexpected trends that we are seeing as a function of metallicity, particularly at high metallicity (see Section~\ref{sec:alpha}).

    \item We present the abundance patterns of Ne/O, S/O, Cl/O, and Ar/O in concert with O/H. We find that these abundance patterns show a constant behavior with metallicity as expected for nucleosynthesis at $\sim$ 7.0 $<$ 12 $+$ log(O/H) $< 8.5$. We confirm the high values of Ne/O at high O/H might be due to inaccurate ICFs \citep[e.g.,][]{izotov06, guseva11, amayo21}. For Ar/O, we find a significant trend with metallicity which is not expected at 12 $+$ log(O/H) $>8.2$. It is possibly explained by the uncertain performance of the ICF of Ar at high metallicities, in particular, when [\ion{Ar}{3}] line alone is used to calculate Ar/O.  

   \item We analyze the abundance trends of Ne/O, S/O, Cl/O, and Ar/O with respect to the relative abundances of Ne, S, Cl, and Ar. Here, we also confirm the trend at high metallicities as the Ne/H abundance increases. The explanation is still unclear, and a future exploration using chemical evolution models \citep{ kobayashi20, alexander23, amayo21} is needed. In a similar way, the Ar/O \textit{vs.}\ Ar/H produces a significant trend as Ar/H increases, although the direction of this trend is opposite for CLASSY when compared to the sample of \citet{mirandaperez23}. Again, this suggests an important bias that might be related to the ICFs of Ar. Therefore, we stress the importance of revising the construction of the ICFs for all the elements, specially when they are applied to galaxies whose spectra are integrated. 
   
    \item We use the CLASSY abundance patterns as a reference to compare with the results of Ne, S, and Ar in z $> 4$ galaxies observed with the $JWST$. We carefully selected those galaxies with direct measurement of the metallicity using the \Te-sensitive method.  We confirm that galaxies at $z> 7$ show similar values of Ne/O to $z\sim0$ galaxies. However, a group of high-$z$ galaxies show very low values of Ne/O \citep[see also,][]{isobe23a}. For S/O and Ar/O, we selected galaxies where the ionic abundances were calculated using a fixed value of \Te and $n_{\rm e}$. However, we stress that fixing a unique value of these two physical properties can result in a significant correlation with O/H (see Figure~\ref{fig:high-z-comp}). Therefore, in this analysis, we confirm that there is no evolution of Ne/O, S/O, and Ar/O across cosmic time and that S/O and Ar/O should be robustly evaluated in $z > 4$ SFGs. 

    \item The analysis of the MZR and FMR derived in CLASSY and $z > 4$ galaxies show no redshift evolution in terms of scatter. For the MZR, CLASSY galaxies show a steeper slope than those studied at high$-z$ and a probable evolution at $z>8$ \citep[e.g.,][]{nakajima23}. Note that we consider here those SFGs with abundance derived using the \Te-sensitive method.  We present a new set of scaling relations based on CLASSY galaxies for Ne, S, Cl, and Ar (see Appendix~\ref{apped:galaxy-properties-elementes}), which shows a slope similar to O, which is expected for the nucleosynthesis involved in those elements. We also report the M-NeR relation for high$-z$ galaxies, whose slope is steeper than the CLASSY galaxies and shifted to lower values of Ne.  
\end{itemize}

Finally, the measurements of the physical conditions and chemical abundance of CLASSY allowed us to assess important biases involved in chemical abundance determinations of different elements. Therefore, CLASSY can be used as a reference to the abundance patterns of $z\sim0$ and high-$z$ SFGs. Particularly, there are different \textit{JWST} programs dedicated to the analysis of the chemical compositions of galaxies across cosmic time. These programs are focused on the detection of \Te-sensitive emission lines at $z\sim3$ \citep[CECILIA;][]{strom23}, Aurora (\textit{JWST} program ID:1914), and EXCELS (\textit{JWST} program ID: 3543) at $z=2-5$. In addition, this analysis will be useful 
for the upcoming observations of the \textit{Extremely Large Telescopes} that will need robust abundance templates at $z\sim0$ to trace and interpret the chemical enrichment of the Universe.  We stress that the selection of ICFs plays an essential role in chemical abundance determinations of metals; however, much work remains to address the biases involved with the building of ICFs to correctly represent the broad range of conditions seen in SFGs. Our analysis with CLASSY provides a careful assessment of the abundance ratios of Ne, S, Cl, and Ar that can help to constrain the nucleosynthesis of chemical evolution models for a broad range of metallicities.


 The CLASSY team thanks the referee for thoughtful feedback that improved this paper. We thank Chiaki Kobayashi for discussions on the chemical evolution of $\alpha$-elements.
 KZA-C and DAB are grateful for the support for this program, HST-GO-15840, that was provided by 
NASA through a grant from the Space Telescope Science Institute, which is operated by the Associations of Universities for Research in Astronomy, 
Incorporated, under NASA contract NAS5-26555. 
The CLASSY collaboration extends special gratitude to the Lorentz Center for useful discussions 
during the "Characterizing Galaxies with Spectroscopy with a view for JWST" 2017 workshop that led 
to the formation of the CLASSY collaboration and survey.
The CLASSY collaboration thanks the COS team for all their assistance and advice in the 
reduction of the COS data. 
KZA-C and FC acknowledge support from a UKRI Frontier Research Guarantee Grant (PI Cullen; grant reference EP/X021025/1).
BLJ and MM are thankful for support from the European Space Agency (ESA).
JB acknowledges support by Fundação para a Ciência e a Tecnologia (FCT) through the research grants UIDB/04434/2020 and UIDP/04434/2020, through work contract No. 2020.03379.CEECIND, and through FCT project PTDC/FISAST/4862/2020. RA acknowledges support from ANID Fondecyt Regular 1202007. 
This work also uses observations obtained with the Large Binocular Telescope (LBT). The LBT is an international collaboration among institutions in the United States, Italy and Germany. LBT Corporation partners are: The University of Arizona on behalf of the Arizona Board of Regents; Istituto Nazionale di Astrofisica, Italy; LBT Beteiligungsge-sellschaft, Germany, representing the Max-Planck Society,The Leibniz Institute for Astrophysics Potsdam, and Heidelberg University; The Ohio State University, University of
19 Notre Dame, University of Minnesota, and University of Virginia.

Funding for SDSS-III has been provided by the Alfred P. Sloan Foundation, the Participating Institutions, the National Science Foundation, and the U.S. Department of Energy Office of Science. 

The SDSS-III website is http://www.sdss3.org/.
SDSS-III is managed by the Astrophysical Research Consortium for the Participating Institutions of the SDSS-III Collaboration including the University of Arizona, the Brazilian Participation Group, Brookhaven National Laboratory, Carnegie Mellon University, University of Florida, the French Participation Group, the German Participation Group, Harvard University, the Instituto de Astrofisica de Canarias, the Michigan State/Notre Dame/JINA Participation Group, Johns Hopkins University, Lawrence Berkeley National Laboratory, Max Planck Institute for Astrophysics, Max Planck Institute for Extraterrestrial Physics, New Mexico State University, New York University, Ohio State University, Pennsylvania State University, University of Portsmouth, Princeton University, the Spanish Participation Group, University of Tokyo, University of Utah, Vanderbilt University, University of Virginia, University of Washington, and Yale University.

This work also uses the services of the ESO Science Archive Facility,
observations collected at the European Southern Observatory under 
ESO programmes 096.B-0690, 0103.B-0531, 0103.D-0705, and 0104.D-0503.


\software{ \texttt{jupyter} \citep{kluyver16}, \texttt{astropy} \citep{astropy:2018, astropy:2022}, {\tt PyNeb} \citep{luridiana15}, {\tt LinRegConf} \citep{Flury_LinRegConf_2024}.}

\bibliography{mybib}

\appendix

\section{Temperature structure of CLASSY}
\label{appen:temperatures}

\begin{figure*}
\begin{center}
    \includegraphics[width=0.80\textwidth, trim=30 0 30 0,  clip=yes]{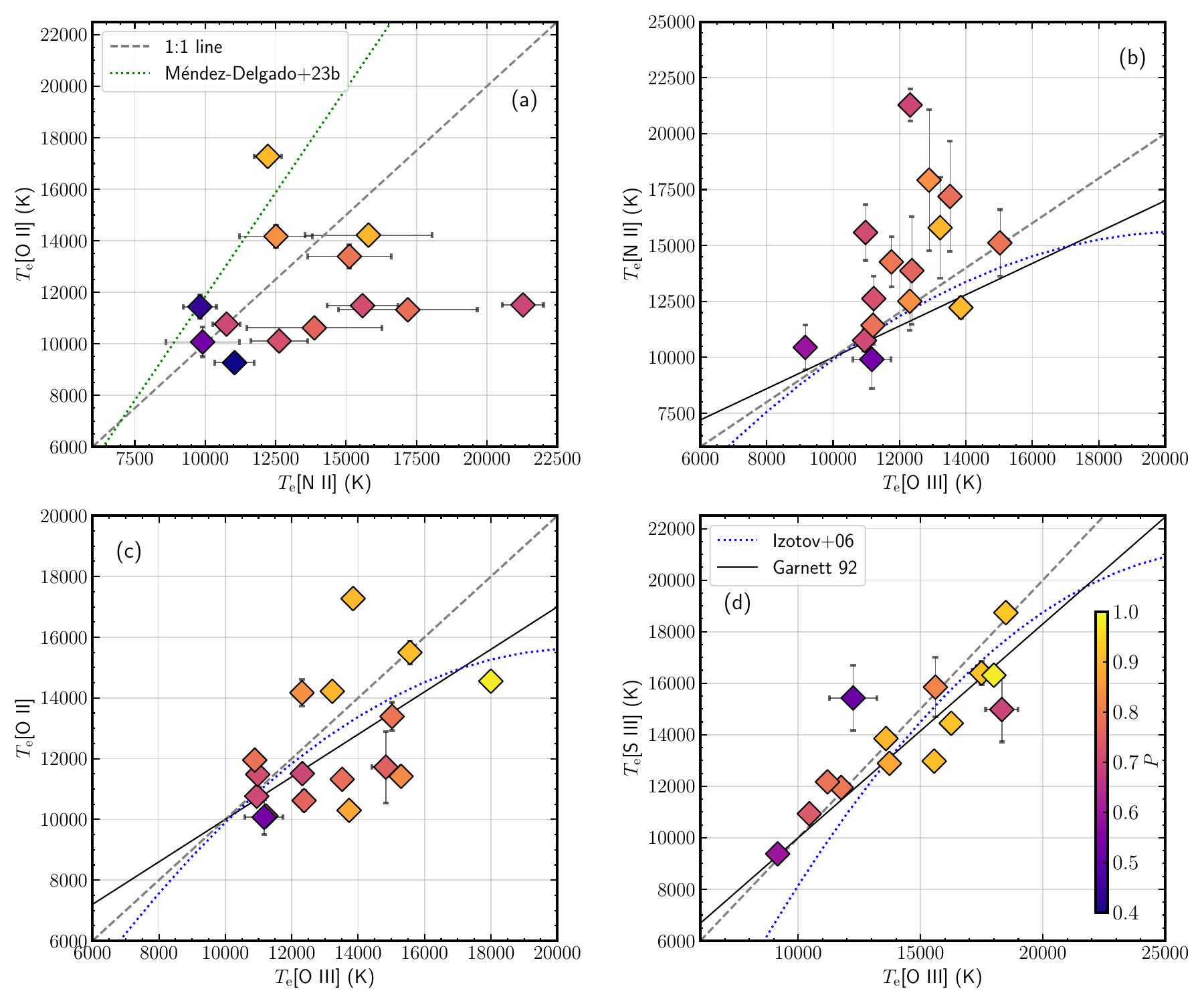}
        \caption{Temperature relations for the CLASSY sample as a function of the ionization parameter ($P$ = [\ion{O}{3}] \W\W4959,5007/([\ion{O}{2}] \W3727 + [\ion{O}{3}] \W\W4959,5007)). \textit{Top:} The low-ionization \To\- \TN\, relation and the \TN-\TO\ relation representative of the low and high ionization gas. \textit{Bottom}: A comparison between the \To-\TO\ relation, and  
        the intermediate and high ionization zones, \TS\ and \TO. The dashed line represents the one-to-one relation. The temperature relations of \citet{garnett92},  \citet{izotov06}, \citet{Rogers21}, and \citet{mendez-delgado23b} are represented in different lines and are labeled in each panel, respectively. While \To\ and \TS as a function of \TO\ follow the temperature relation from the literature, the \TN-\TO/\To relations show a large dispersion with a significant depart from those relations of the literature due to at high values of \TN. }
\label{fig:t-t-relations}
\end{center}
\end{figure*}

 In Figure~\ref{fig:t-t-relations} we present the temperature relations implied on the results of \To, \TN, \TS, and \TO\ for CLASSY color coded with the ionization parameter, $P$ = [\ion{O}{3}] \W\W 4959, 5007/([\ion{O}{2}] \W3727 + [\ion{O}{3}] \W\W 4959, 5007). For reference, the dashed line represents the one-to-one relation in each plot. The different lines in Figure~\ref{fig:t-t-relations} represent some common temperature relations used in the literature \citep[e.g.,][]{garnett92, izotov06}. For a sample of 13 CLASSY galaxies, we analyze the \TN\ -\To\ relation (see panel (a) of Figure~\ref{fig:t-t-relations}), which trace the low ionization zone of the nebula. Some galaxies show significant departures to high values of \TN\ (see Table~\ref{tab:physical_conditions}). We have derived \TN\ up to 6000 K larger than \To, which is not expected if \TN\ and \To\ trace similar gas conditions. For example, for J0036-3333, we measured \TN\, = 27200$\pm$100 K, while a higher value was reported by \citet[][\TN\ = 52700 K]{menacho21} for a similar region than in these observations. Another example is J0127-0619, with a value of \TN\, = 30400$\pm$2700 (see Table~\ref{tab:physical_conditions}). This galaxy shows a broad component on the \Te-sensitive [\ion{N}{2}] \W5755 line and the high value of \TN\ could be related to the contribution of that component \citep[see also][]{james09}. However, we have used the fluxes related to the narrow component in our calculations (see also \PIV). 

One of the explanations for those high values in \TN\ could be associated with the contribution by recombination. [\ion{N}{2}] \W5755 might be affected for the contribution of recombination process of N$^{2+}$/H$^{+}$ \citep{rubin86} overestimating the value of \TN.  Typically, to estimate the contribution by recombination to [\ion{N}{2}] \W5755, we use the expression derive by \citet{Liu00}, which depends on the measurement of N$^{2+}$/H$^{+}$ and can be estimated as N$^{2+}$/H$^{+}$ = N/H - N$^{+}$/H$^{+}$. Note that such an estimate is quite uncertain since it depends on the estimate of the N$^{2+}$/H$^{+}$ abundance. 
\citet{menacho21} also analyzed the contribution by recombination to the [\ion{N}{2}] \W5755 line \citep[][]{Liu00, stasinska05}. Those authors found that this process is about 50\%, which implies an overestimate of \TN. However, the [\ion{O}{2}] \W\W 7220,30 lines are even more affected by recombination; these authors found a minor contribution by this process to the emission of [\ion{O}{2}] \W\W 7220,30 (less than 5\%). Since both auroral lines should be affected in the same way, it is still uncertain as to what is driving the high values of \TN\ \citep[]{menacho21, Loaiza-Agudelo20}.  In addition, in panel (a) of Figure~\ref{fig:t-t-relations} we have added the temperature relation of \citet{mendez-delgado23b} based on a sample of \hii\ regions, which shows a steeper relation in comparison to CLASSY and the 1:1 line. In general, our results of the \To-\TN\ relation are more consistent with a 1:1 line \citep[see also][]{Rogers22}. 

Panels (b) and (c) of Figure~\ref{fig:t-t-relations}, show the temperature relations of \TO\ as a function of \TN\ and \To, respectively.  We found a large discrepancy between \TN\, $\emph{vs.}$ \TO\, due to the high values of \TN\ for this sample. The dispersion in the \TN-\TO\ relationship is discussed in previous studies, mainly for samples of \hii~regions \citep[e.g.][]{pilyugin07, croxall16, arellano-cordova20, berg20, Rogers21, mendez-delgado+23a}. Although part of the dispersion might be real, these authors claim that part of such dispersion could be due to dependence on the ionization parameter (see Figure~\ref{fig:t-t-relations}), the difference of the age of the stellar population and observational problems such as the detection and measurement of faint \Te-sensitive lines. 

For the  \To\, $\emph{vs.}$ \TO\ relation, we found that SFGs with $P < 0.8$ follow the relations of \citet{garnett92} and \citet{izotov06} for the range of temperatures associated to the CLASSY galaxies, while that that some galaxies with $P >0.8$ departs to high values of \To. 
Recently, \citet{mendez-delgado23b} analyzed the density and temperature structure in \hii\ regions, finding that the presence of high density clouds can have a direct impact on \To\ in comparison to \TN. For those galaxies in our sample with measurements of \To\ and \TN, we have recalculated O$^{+}$ to evaluate the impact of \To\ and \TN\ on the total abundance of O. We can perform this exercise for 11 galaxies with measurements of \To, \TN\ and \TO. For these galaxies, we compare O/H implied by the contribution of O$^{+}$ by comparing the results of \To\ and \TN. We find that the differences in O/H are lower than 0.09 dex for most of the galaxies, while for four galaxies (J0021+0052, J1359+572, J1025+3622, J1025+3622, J1105+4444) the differences are in the range of 0.12-0.17 dex. In particular, for such galaxies, the values of \TN\ are relatively high (see also Figure~\ref{fig:t-t-relations}). As an alternative, we also analyzed the results implied using a temperature relation to estimate \To\ from \TO. Thus, we compare the differences in O/H  derived using the temperature relation and the direct measurement of \To. We found differences lower than 0.08 dex. Therefore, it is reasonable to use the direct measurement of \To\ to calculate O$^{+}$ for CLASSY.

In panel (c) of Figure~\ref{fig:t-t-relations}, we show the \TS\ $\emph{vs.}$ \TO\ relationship, which represents \Te\ associated to the intermediate and high ionization zones, $T_{\rm e}$(Int.) and $T_{\rm e}$(High), respectively. We find that most of the CLASSY galaxies follow the trend implied by the relation of \cite{garnett92}, as was reported previously in \PIV\ and \PV. 

As a summary, from the different temperature diagnostics derived in CLASSY, we selected $T_{\rm e}$[\ion{O}{2}] as the preferred $T_{\rm e}$(Low) when \To\, is measured, otherwise we use the value derived for \TN\, with exception of those galaxies with very high \TN. For such galaxies, we use as reference the measurement of $T_{\rm e}$[\ion{S}{3}] or $T_{\rm e}$[\ion{O}{3}] to estimate $T_{\rm e}$(Low). For J0808+3948, we use \TN\ as $T_{\rm e}$(Low).

\section{Galaxy Property correlations}
\label{apped:galaxy-properties-elementes}
\begin{figure*}
\begin{center}
    \includegraphics[width=0.9\textwidth, trim=30 0 30 0,  clip=yes]{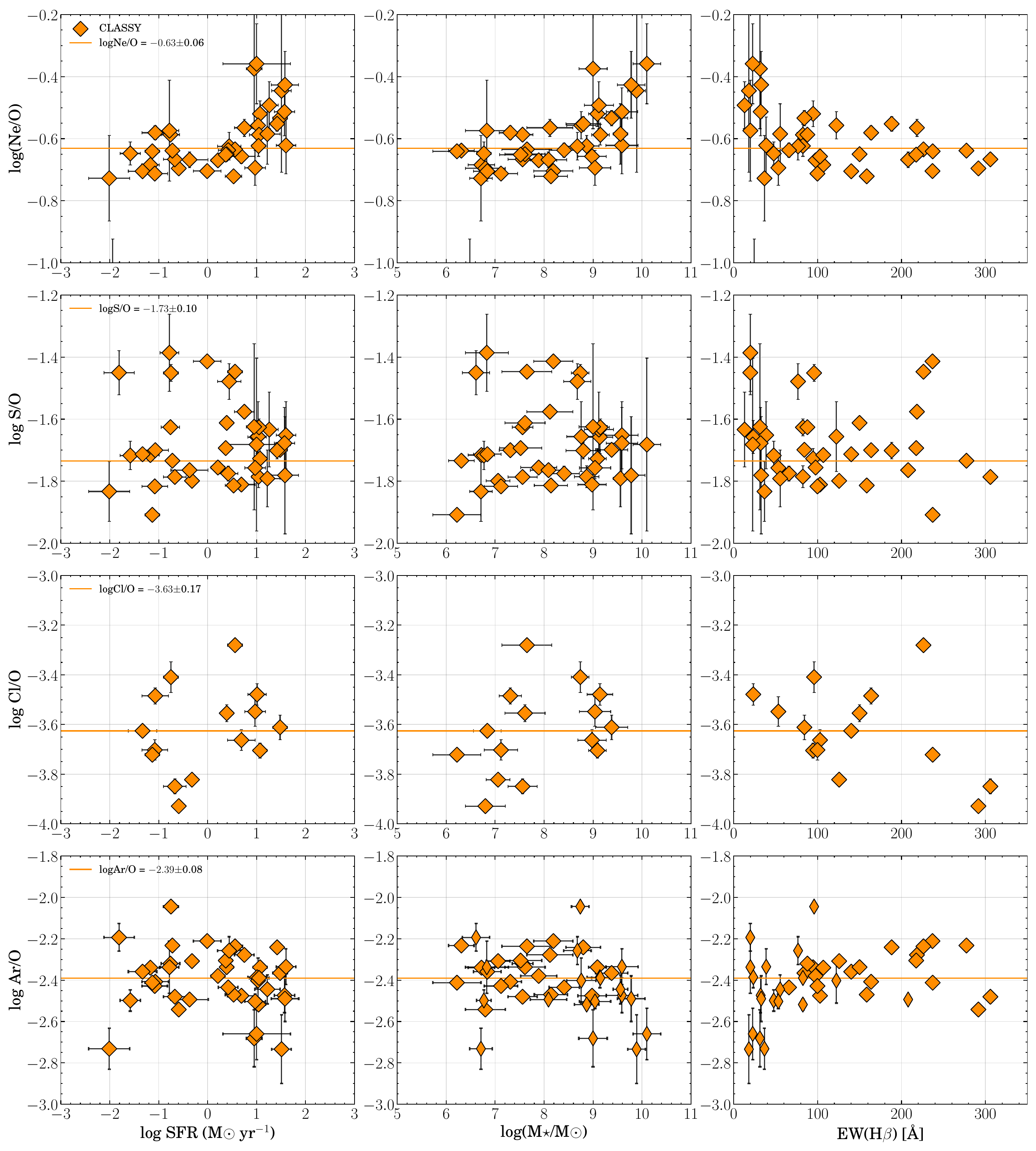}
        \caption{The abundance ratios of Ne/O, S/O, Cl/O, and Ar/O with respect to SFR (left), stellar mass (middle), and EW(H$\beta$) (right) for CLASSY. The orange line indicates the abundance ratio for CLASSY (see also Figure~\ref{fig:alpha_elements}).  The comparison of the abundance ratios derived in SFGs of the CLASSY sample shows a constant trend with the galaxy properties. The Cl/O abundance ratio shows a tentative correlation with stellar mass. A similar behavior is also shown in Figure~\ref{fig:alpha_elements} as a function of O/H.}
\label{apped:ratios-properties}
\end{center}
\end{figure*}

\begin{figure}
\begin{center}
    \includegraphics[width=0.7\textwidth, trim=15 0 15 0,  clip=yes]{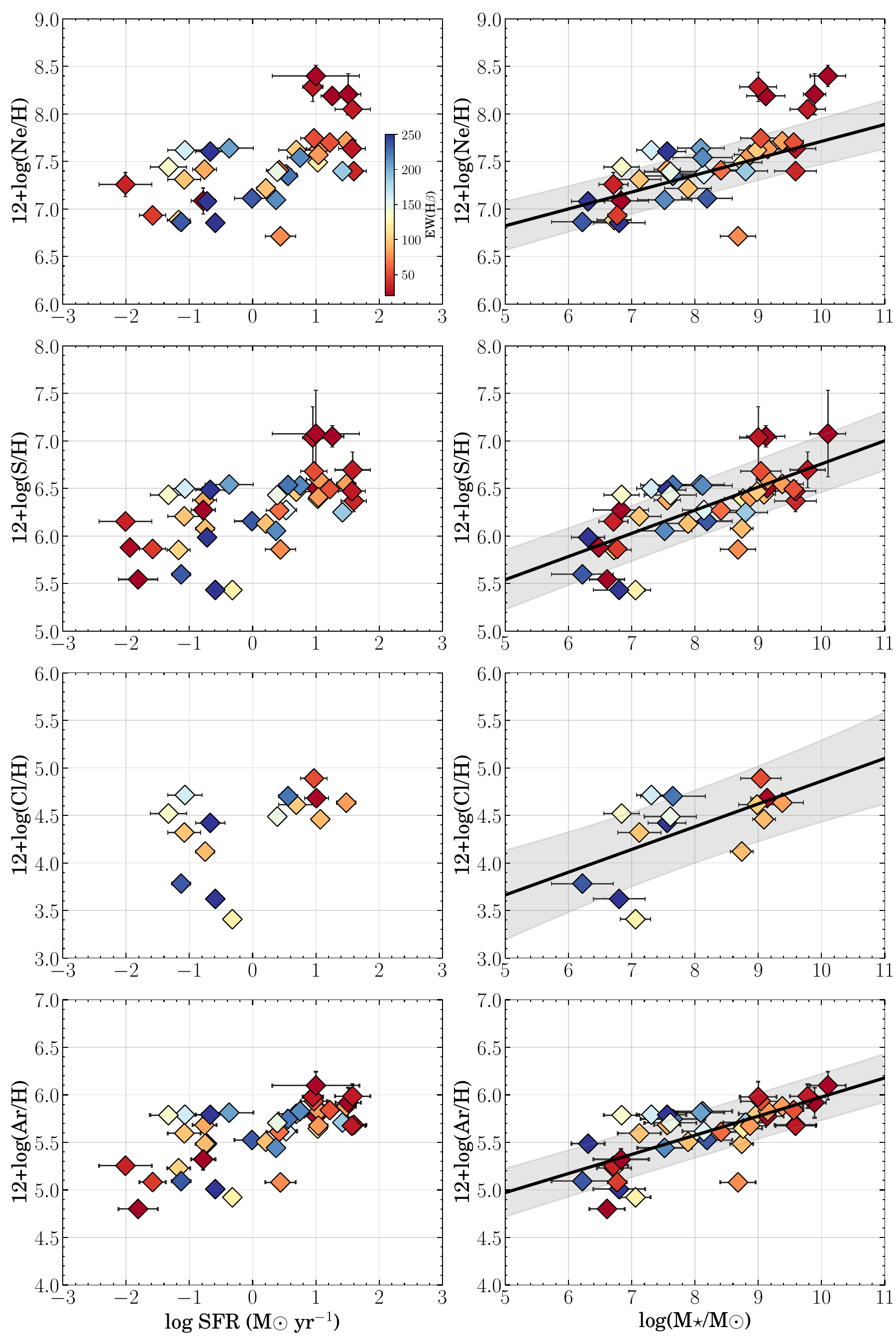}
        \caption{The total abundance ratios of Ne, S, Cl, and Ar with respect to SFR (left) and the stellar mass (right) for CLASSY in color codded with the EW of H$\beta$. Note that relation between Ne/H and MNeR is the same reported in Fig.~\ref{fig:mzr} for CLASSY. The solid lines show the best fit to the data with the stellar mass (see Table~\ref{tab:abundance_ratios}). The comparison with the SFR shows constant behaviors at log(SFR) $<0$. A tentative correlation of S/H and Ar/H with respect to SFR could be evident, but lower values of SFR show a large scatter as the abundance patterns decrease, showing no significant correlation with the SFR.  }
\label{fig:total-properties}
\end{center}
\end{figure}

For this section, we investigate any correlation between Ne/O, S/O, Cl/O, and Ar/O ratios (and Ne/H, S/H, Cl/H, Ar/H) and the galaxy properties such as the stellar mass, SFR, and EW(H$\beta$). We use the galaxy properties of CLASSY derived in \PI.
Figure~\ref{apped:ratios-properties} illustrates the comparison between Ne/O, S/O, Cl/O, and Ar/O and SFR (left), Stellar mass (middle), and EW(H$\beta$) (right), respectively. In general, all the abundance ratios show a constant trend with all the galaxy properties. For Ne/O, there is a slight correlation between the SFR and the stellar mass as Ne/O increases. As we mentioned in Sec.~\ref{sec:neon}, such a slight trend could be due to issues with the ICF of Ne.  For Cl/O, the dispersion is larger than the rest of the elements showing a possible correlation with the stellar mass and EW(H$\beta$). One possibility to confirm or discard the trend of Cl/O with the galaxy properties is the detection of  [\ion{Cl}{3}] \W\W 5518,31 at higher matllicities or log(M$_{\star}$/M$\odot$) $> 10$. However, [\ion{Cl}{3}] \W\W 5518,31 are very faint making difficult the analysis of the Cl/O ratios. However, our results with CLASSY in comparison with \hii\ regions with CHAOS suggest a possible correlation of Cl/O between metallicity and the stellar mass (see also Figure~\ref{fig:alpha_elements}).
In addition, Figure~\ref{fig:total-properties} shows the correlations between  Ne/H, S/H, Cl/H, and Ar/H with respect to the SFR (left) and the stellar mass (right). Overall, such elements show a trend with those galaxy properties. Although, for Cl, the dispersion is larger it is also evident a slight trend with SFR and the stellar mass. Also note the total abundance derived for each element depends on the measurement of O/H (e.g., Ne/H =Ne/O $\times$ O/H ). 
We have fitted a linear relation to the results of Figure~\ref{fig:total-properties}, which are presented in Table~\ref{tab:relationships-properties}. In general, we found slopes ranging from 0.20 to 0.24. The stepper slopes correspond to S and Cl, 0.24$\pm$0.04 and 0.24$\pm$0.09, respectively. However, we calculated a large scatter of the fit with respect to the data of 0.34 for Cl. For Ne and Ar, we found similar values of the slope with a scatter of $\sigma \sim0.20$.(see Table~\ref{tab:relationships-properties}). Note that for Ne, we discarded the galaxies with high values of Ne/O to the fit. 

For the comparison with the SFR, the different abundance patterns show a large dispersion at low values of SFR $< 0$ and a constant behavior, 
while a tentative correlation with less scatter is more evident when the SFR increases for S and Ar. Although there is not a correlation between the EW(H$\beta$), Ne/H and S/H increase at fixed SFR in galaxies with low values of EW(H$\beta$). However, the large dispersion at lower values of Ne, S, and Ar is consistent with no correlation of these elements as SFR decreases.

\section{Tables: Physical conditions, chemical abundances and galaxy properties}\label{appen:tables}

In Table~\ref{tab:physical_conditions}, we present the results of \Te\ and \Ne\ for CLASSY, and the selected temperature structure in columns 7-9 to determine the chemical abundances (see Section~\ref{sec:physical conditions} and Appendix \ref{appen:temperatures}). 
In Table~\ref{tab:abundance_ratios} list the stellar mass and SFR of CLASSY in columns 2-4. The metallicity, and the abundance ratios of Ne/O, S/O, and Cl/O and Ar/O in columns 4-8.

\input{temden} 
\input{Sample_provicional} 

\end{document}

%% file: authors.tex
\author[0000-0002-2644-3518]{Karla Z. Arellano-C\'{o}rdova}
\affiliation{Institute for Astronomy, University of Edinburgh, Royal Observatory, Edinburgh, EH9 3HJ, UK}
\affiliation{Department of Astronomy, The University of Texas at Austin, 2515 Speedway, Stop C1400, Austin, TX 78712, USA}

\author[0000-0002-4153-053X]{Danielle A. Berg}
\affiliation{Department of Astronomy, The University of Texas at Austin, 2515 Speedway, Stop C1400, Austin, TX 78712, USA}

\author[0000-0003-2589-762X]{Matilde Mingozzi}
\affiliation{AURA for ESA, Space Telescope Science Institute, 3700 San Martin Drive, Baltimore, MD 21218, USA}

\author[0000-0003-4372-2006]{Bethan L. James}
\affiliation{AURA for ESA, Space Telescope Science Institute, 3700 San Martin Drive, Baltimore, MD 21218, USA}

\author[0000-0000-0000-0000]{Noah S. J., Rogers}
\affiliation{Center for Interdisciplinary Exploration and Research in Astrophysics (CIERA), Northwestern University, 1800 Sherman Ave., Evanston, IL, 60201, USA}

\author[0000-0003-0605-8732]{Evan D. Skillman}
\affiliation{Minnesota Institute for Astrophysics, University of Minnesota, 116 Church Street SE, Minneapolis, MN 55455, USA}

\author[0000-0002-3736-476X]{Fergus, Cullen}
\affiliation{Institute for Astronomy, University of Edinburgh, Royal Observatory, Edinburgh, EH9 3HJ, UK}

\author[0009-0006-2621-6979]{Ryan, Alexander}
\affiliation{E.A. Milne Centre for Astrophysics, University of Hull, Hull, HU6 7RX, UK}

\author[0000-0001-5758-1000]{Ricardo O. Amor\'{i}n}
\affiliation{Instituto de Investigaci\'{o}n Multidisciplinar en Ciencia y Tecnolog\'{i}a, Universidad de La Serena, Raul Bitr\'{a}n 1305, La Serena 2204000, Chile}
\affiliation{Departamento de Astronom\'{i}a, Universidad de La Serena, Av. Juan Cisternas 1200 Norte, La Serena 1720236, Chile}

\author[0000-0002-0302-2577]{John Chisholm}
\affiliation{Department of Astronomy, The University of Texas at Austin, 2515 Speedway, Stop C1400, Austin, TX 78712, USA}

\author[0000-0001-8587-218X]{Matthew Hayes}
\affiliation{Stockholm University, Department of Astronomy and Oskar Klein Centre for Cosmoparticle Physics, AlbaNova University Centre, SE-10691, Stockholm, Sweden}

\author[0000-0003-1127-7497]{Timothy Heckman}
\affiliation{Center for Astrophysical Sciences, Department of Physics \& Astronomy, Johns Hopkins University, Baltimore, MD 21218, USA}

\author[0000-0003-4857-8699]{Svea Hernandez}
\affiliation{AURA for ESA, Space Telescope Science Institute, 3700 San Martin Drive, Baltimore, MD 21218, USA}

\author[0000-0002-5320-2568]{Nimisha Kumari}
\affiliation{AURA for ESA, Space Telescope Science Institute, 3700 San Martin Drive, Baltimore, MD 21218, USA}

\author[0000-0003-2685-4488]{Claus Leitherer}
\affiliation{Space Telescope Science Institute, 3700 San Martin Drive, Baltimore, MD 21218, USA}

\author[0000-0001-9189-7818]{Crystal L. Martin}
\affiliation{Department of Physics, University of California, Santa Barbara, Santa Barbara, CA 93106, USA}

\author[0000-0000-0000-0000]{Michael Maseda}
\affiliation{Department of Astronomy, University of Wisconsin-Madison, Madison, WI 53706, USA}

\author[0000-0003-2804-0648]{Themiya Nanayakkara}
\affiliation{Swinburne University of Technology, Melbourne, Victoria, AU}

\author[0000-0000-0000-00000]{Kaelee Parker}
\affiliation{Department of Astronomy, The University of Texas at Austin, 2515 Speedway, Stop C1400, Austin, TX 78712, USA}

\author[0000-0002-5269-6527]{Swara Ravindranath}
\affiliation{Astrophysics Science Division, NASA Goddard Space Flight Center, 8800 Greenbelt Road, Greenbelt, MD 20771, USA}
\affiliation{Center for Research and Exploration in Space Science and Technology II, Department of Physics, Catholic University of America, 620 Michigan Ave N.E., Washington DC 20064, USA}

\author[0000-0001-6369-1636]{Allison L. Strom}
\affiliation{Department of Astrophysical Sciences, 4 Ivy Lane, Princeton University, Princeton, NJ 08544, USA}

\author[0000-0002-0743-9994]{Fiorenzo Vincenzo}
\affiliation{E.A. Milne Centre for Astrophysics, University of Hull, Hull, HU6 7RX, UK}

\author[0000-0001-8289-3428]{Aida Wofford}
\affiliation{Instituto de Astronom\'{i}a, Universidad Nacional Aut\'{o}noma de M\'{e}xico, Unidad Acad\'{e}mica en Ensenada, Km 103 Carr. Tijuana-Ensenada, Ensenada 22860, M\'{e}xico}
\affiliation{Department of Astronomy \& Astrophysics, University of California San Diego, 9500 Gilman Drive, La Jolla, CA 92093, USA}

%% file: atomic_data.tex
 \begin{table*}\footnotesize
 \caption{Atomic data used in this work}
 \begin{center}
 \begin{tabular}{lcccc}
 \hline
 \multicolumn{1}{l}{Ion} & \multicolumn{1}{c}{Transition Probabilities (A$_{ij}$)} & \multicolumn{1}{c}{A$_{ij}$$^\star$} & \multicolumn{1}{c}{Collision Strengths ($\Upsilon_{ij}$)}  &  \multicolumn{1}{c}{$\Upsilon_{ij}$$^\star$}\\
 \hline

O$^{+}$   &  \citet{fft04}  & FFT04  & \citet{Kisielius:2009} & Kal09\\
O$^{2+}$  &  \citet{fft04}  & FFT04  & \citet{aggarwal99} & AK99\\
 N$^{+}$   &  \citet{fft04} & FFT04  & \citet{tayal11}   & T11\\
Ne$^{2+}$  &  \citet{McLaughlin:2011} & McL11 & \citet{McLaughlin:2011} & McL11\\
S$^{+}$   &  \citet{Podobedova:2009} & PKW09 &  \citet{Tayal:2010} & TZ10\\
S$^{2+}$  &  \citet{Podobedova:2009} & PKW09 &  \citet{Grieve:2014} & GRHK14\\

Cl$^{2+}$ &  \citet{Fritzsche:1999} &  Fal99  & \citet{Butler:1989} & BZ89\\
Cl$^{3+}$ &  \citet{Kaufman:1986},  & KS86-MZ82-EM84 & \citet{Galavis:1995} & GMZ95\\
          &  \citet{Mendoza:1982a}, \citet{Ellis:1984} &                     & \\

Ar$^{2+}$ &   \citet{Mendoza:1983}, \citet{Kaufman:1986} &  M83-KS86 & \citet*{Galavis:1995} & GMZ95\\
Ar$^{3+}$ &   \citet{Mendoza:1982b} & MZ82 & \citet{Ramsbottom:1997} & RB97\\

 \hline
 \end{tabular}
 \end{center}
 \tablecomments{$^\star$ Identified as in {\sc Pyneb:} $\Upsilon_{ij}$\_coll and "A$_{ij}$\_atom", respectively. } 
 \label{tab:atomic_data}
 \end{table*}

 

%% file: temden.tex
\begin{deluxetable*}{cccccc|ccc}
\centering
\tablewidth{0pt}
\caption{Electron density and temperature structure of CLASSY}
\tablehead{          & $n_{e}$[S II] & $T_{\rm e}$[O II]& $T_{\rm e}$[N II] & $T_{\rm e}$[S III] & $T_{\rm e}$[O III] & $T_{e}$(Low) & $T_{e}$(Int.) & $T_{e}$(High) \\
Galaxy & (cm$^{-3}$) & (K) & (K) & (K) & (K) & (K) & (K) & (K)
}
\startdata
  J0021+0052  &    97$\pm$32  &  11500$\pm$ 300  &  21300$\pm$ 800  &      \nodata     &  12300$\pm$ 300  &  11500$\pm$ 300  &  11900$\pm$ 200  &  12300$\pm$ 300  \\
  J0036-3333  &   $<$100  &      \nodata     &  27200$\pm$ 100  &  11400$\pm$ 100  &      \nodata     &  10600$\pm$ 400  &  11400$\pm$ 900  &  11700$\pm$ 300  \\
  J0127-0619  &   408$\pm$40  &      \nodata     &  30400$\pm$2700  &      \nodata     &  18300$\pm$ 400  &  15800$\pm$1700  &  16900$\pm$ 400  &  18300$\pm$ 400  \\
  J0337-0502  &   180$\pm$10  &      \nodata     &      \nodata     &  20200$\pm$ 100  &      \nodata     &  16500$\pm$ 600  &  20200$\pm$1500  &  22200$\pm$ 100  \\
  J0405-3648  &    28$\pm$16  &      \nodata     &      \nodata     &  23500$\pm$1800  &      \nodata     &  18800$\pm$1400  &  23500$\pm$2800  &  26300$\pm$ 100  \\
  J0808+3948  &  1179$\pm$100  &      \nodata     &   7500$\pm$ 100  &      \nodata     &      \nodata     &   7500$\pm$ 100  &   6900$\pm$1000  &   6500$\pm$ 100  \\
  J0823+2806  &   144$\pm$24  &  10800$\pm$ 100  &  10800$\pm$ 500  &      \nodata     &  10900$\pm$ 100  &  10800$\pm$ 100  &  10800$\pm$ 100  &  10900$\pm$ 100  \\
  J0926+4427  &    96$\pm$70  &  11700$\pm$1100  &      \nodata     &      \nodata     &  14800$\pm$ 400  &  11700$\pm$1100  &  14000$\pm$ 400  &  14800$\pm$ 400  \\
  J0934+5514  &   $<$100  &      \nodata     &      \nodata     &      \nodata     &  20400$\pm$ 200  &  17300$\pm$1800  &  18600$\pm$ 200  &  20400$\pm$ 200  \\
  J0938+5428  &   106$\pm$37  &  11500$\pm$ 200  &  15600$\pm$1200  &      \nodata     &  11000$\pm$ 200  &  11500$\pm$ 200  &  10800$\pm$ 200  &  11000$\pm$ 200  \\
  J0940+2935  &     9$\pm$16  &      \nodata     &      \nodata     &  15400$\pm$1300  &  12200$\pm$1000  &  11600$\pm$1400  &  15400$\pm$1300  &  12200$\pm$1000  \\
  J0942+3547  &    50$\pm$18  &      \nodata     &  17900$\pm$3000  &      \nodata     &  12900$\pm$ 100  &  12000$\pm$1300  &  12400$\pm$ 100  &  12900$\pm$ 100  \\
  J0944+3442  &   113$\pm$48  &      \nodata     &      \nodata     &      \nodata     &  15300$\pm$1600  &  13700$\pm$1800  &  14400$\pm$1300  &  15300$\pm$1600  \\
  J0944-0038  &   138$\pm$56  &  15500$\pm$ 400  &      \nodata     &  14800$\pm$ 500  &  15600$\pm$ 100  &  15500$\pm$ 400  &  14800$\pm$ 500  &  15600$\pm$ 100  \\
  J1016+3754  &    36$\pm$25  &      \nodata     &      \nodata     &  16400$\pm$ 500  &  17500$\pm$ 200  &  15200$\pm$1600  &  16400$\pm$ 500  &  17500$\pm$ 200  \\
  J1024+0524  &    76$\pm$19  &  11400$\pm$ 200  &      \nodata     &      \nodata     &  15300$\pm$ 100  &  11400$\pm$ 200  &  14400$\pm$ 100  &  15300$\pm$ 100  \\
  J1025+3622  &   198$\pm$56  &  10600$\pm$ 300  &  13900$\pm$2600  &      \nodata     &  12400$\pm$ 200  &  10600$\pm$ 300  &  12000$\pm$ 200  &  12400$\pm$ 200  \\
  J1044+0353  &   267$\pm$19  &      \nodata     &      \nodata     &  23400$\pm$ 200  &  17600$\pm$ 200  &  15300$\pm$1600  &  23400$\pm$ 200  &  17600$\pm$ 200  \\
  J1105+4444  &   113$\pm$23  &  10100$\pm$ 100  &  12600$\pm$1100  &      \nodata     &  11200$\pm$ 100  &  10100$\pm$ 100  &  11000$\pm$ 100  &  11200$\pm$ 100  \\
  J1112+5503  &   408$\pm$93  &  11400$\pm$ 400  &   9800$\pm$ 600  &      \nodata     &      \nodata     &  11400$\pm$ 400  &  12600$\pm$1300  &  12100$\pm$ 100  \\
  J1119+5130  &   $<$100  &      \nodata     &      \nodata     &  15800$\pm$1100  &  15600$\pm$ 400  &  13900$\pm$1500  &  15800$\pm$1100  &  15600$\pm$ 400  \\
  J1129+2034  &    83$\pm$15  &      \nodata     &      \nodata     &  10900$\pm$ 100  &  10500$\pm$ 100  &  10300$\pm$1200  &  10900$\pm$ 100  &  10500$\pm$ 100  \\
  J1132+1411  &    95$\pm$24  &  12000$\pm$ 200  &      \nodata     &      \nodata     &  10900$\pm$ 100  &  12000$\pm$ 200  &  10700$\pm$ 100  &  10900$\pm$ 100  \\
  J1132+5722  &   122$\pm$41  &      \nodata     &      \nodata     &  15000$\pm$1300  &  18300$\pm$ 600  &  15800$\pm$1700  &  15000$\pm$1300  &  18300$\pm$ 600  \\
  J1144+4012  &   109$\pm$48  &   8600$\pm$ 500  &      \nodata     &      \nodata     &      \nodata     &   8600$\pm$ 500  &   8500$\pm$1200  &   8100$\pm$ 600  \\
  J1148+2546  &   113$\pm$10  &  10300$\pm$ 100  &      \nodata     &  12900$\pm$ 100  &  13700$\pm$ 100  &  10300$\pm$ 100  &  12900$\pm$ 100  &  13700$\pm$ 100  \\
  J1150+1501  &    85$\pm$12  &      \nodata     &  14300$\pm$1200  &  12000$\pm$ 100  &  11800$\pm$ 100  &  11200$\pm$1200  &  12000$\pm$ 100  &  11800$\pm$ 100  \\
  J1157+3220  &    67$\pm$15  &      \nodata     &  10400$\pm$1000  &   9400$\pm$ 200  &   9200$\pm$ 200  &   9400$\pm$1100  &   9400$\pm$ 200  &   9200$\pm$ 200  \\
  J1200+1343  &   172$\pm$53  &  14200$\pm$ 400  &  12500$\pm$1200  &      \nodata     &  12300$\pm$ 200  &  14200$\pm$ 400  &  11900$\pm$ 200  &  12300$\pm$ 200  \\
  J1225+6109  &    24$\pm$19  &      \nodata     &      \nodata     &  13900$\pm$ 300  &  13600$\pm$ 100  &  12500$\pm$1400  &  13900$\pm$ 300  &  13600$\pm$ 100  \\
  J1253-0312  &   437$\pm$35  &  17300$\pm$ 300  &  12200$\pm$ 500  &      \nodata     &  13800$\pm$ 100  &  17300$\pm$ 300  &  13200$\pm$ 100  &  13800$\pm$ 100  \\
  J1314+3452  &   180$\pm$15  &      \nodata     &  11400$\pm$ 900  &  12200$\pm$ 100  &  11200$\pm$ 100  &  10800$\pm$1200  &  12200$\pm$ 100  &  11200$\pm$ 100  \\
  J1323-0132  &   629$\pm$100  &  14600$\pm$ 300  &      \nodata     &  16300$\pm$ 200  &  18000$\pm$ 100  &  14600$\pm$ 300  &  16300$\pm$ 200  &  18000$\pm$ 100  \\
  J1359+5726  &    69$\pm$34  &  11300$\pm$ 200  &  17200$\pm$2600  &      \nodata     &  13500$\pm$ 200  &  11300$\pm$ 200  &  12900$\pm$ 200  &  13500$\pm$ 200  \\
  J1416+1223  &   295$\pm$61  &  10700$\pm$ 300  &      \nodata     &      \nodata     &      \nodata     &  10700$\pm$ 300  &  11500$\pm$1200  &  11000$\pm$ 200  \\
  J1418+2102  &    73$\pm$13  &      \nodata     &      \nodata     &  18700$\pm$ 100  &  18500$\pm$ 200  &  15900$\pm$1700  &  18700$\pm$ 100  &  18500$\pm$ 200  \\
  J1428+1653  &   119$\pm$62  &  10100$\pm$ 500  &   9900$\pm$1400  &      \nodata     &  11200$\pm$ 600  &  10100$\pm$ 500  &  11000$\pm$ 500  &  11200$\pm$ 600  \\
  J1429+0643  &    63$\pm$50  &  13400$\pm$ 500  &  15100$\pm$1500  &      \nodata     &  15000$\pm$ 200  &  13400$\pm$ 500  &  14200$\pm$ 200  &  15000$\pm$ 200  \\
  J1448-0110  &   120$\pm$29  &  14200$\pm$ 200  &  15800$\pm$2200  &      \nodata     &  13200$\pm$ 100  &  14200$\pm$ 200  &  12700$\pm$ 100  &  13200$\pm$ 100  \\
  J1521+0759  &    76$\pm$68  &   8900$\pm$ 500  &      \nodata     &      \nodata     &      \nodata     &   8900$\pm$ 500  &   8900$\pm$1200  &   8500$\pm$ 100  \\
  J1525+0757  &   186$\pm$60  &   8300$\pm$ 200  &      \nodata     &      \nodata     &      \nodata     &   8300$\pm$ 200  &   8100$\pm$1100  &   7600$\pm$ 100  \\
  J1545+0858  &   153$\pm$10  &      \nodata     &      \nodata     &  14400$\pm$ 100  &  16300$\pm$ 100  &  14400$\pm$1500  &  14400$\pm$ 100  &  16300$\pm$ 100  \\
  J1612+0817  &   484$\pm$88  &   9300$\pm$ 300  &  11000$\pm$ 700  &      \nodata     &      \nodata     &   9300$\pm$ 300  &   9400$\pm$1100  &   9000$\pm$ 100  \\
\enddata  
\tablecomments{In columns 7-9 show the low, intermediate, and high ionization $T_{\rm e}$(low), $T_{\rm e}$(Int), and $T_{\rm e}$(low), respectively, used to calculate the ionic abundances see Section~\ref{sec:physical conditions}.}
\label{tab:physical_conditions}
\end{deluxetable*} 

%% file: Sample_provicional.tex
\begin{deluxetable*}{ccccccccc}
\centering
\setlength{\tabcolsep}{3pt}
\tablewidth{0pt}
\caption{Galaxy properties, metallicity, and chemical abundance ratios of Ne/O, S/O, Cl/O and Ar/O of CLASSY}
\tablehead{Galaxy & $z$ & M$_{star}$ & log SFR & 12+log(O/H)  & log(Ne/O) & log(S/O) & log(Cl/O) & log(Ar/O) \\ }
\startdata
  J0021+0052  &  0.0984  &  $~9.09^{+0.18}_{-0.38}$  &  $~1.07^{+0.14}_{-0.11}$  &  $ 8.16\pm0.17$  &  $-0.52\pm0.03$  &  $-1.73\pm0.03$  &  $-3.70\pm0.03$  &  $-2.34\pm0.03$  \\
  J0036-3333  &  0.0206  &  $~9.14^{+0.26}_{-0.23}$  &  $~1.01^{+0.19}_{-0.21}$  &  $ 8.16\pm0.17$  &      \nodata     &  $-1.66\pm0.06$  &  $-3.48\pm0.06$  &  $-2.39\pm0.05$  \\
  J0127-0619  &  0.0054  &  $~8.74^{+0.18}_{-0.15}$  &  $-0.75^{+0.15}_{-0.13}$  &  $ 7.53\pm0.08$  &      \nodata     &  $-1.45\pm0.03$  &  $-3.41\pm0.06$  &  $-2.04\pm0.02$  \\
  J0337-0502  &  0.0135  &  $~7.06^{+0.24}_{-0.21}$  &  $-0.32^{+0.07}_{-0.11}$  &  $ 7.23\pm0.04$  &      \nodata     &  $-1.80\pm0.01$  &  $-3.82\pm0.02$  &  $-2.31\pm0.01$  \\
  J0405-3648  &  0.0028  &  $~6.61^{+0.28}_{-0.28}$  &  $-1.81^{+0.31}_{-0.27}$  &  $ 6.99\pm0.07$  &      \nodata     &  $-1.45\pm0.07$  &      \nodata     &  $-2.19\pm0.07$  \\
  J0808+3948  &  0.0912  &  $~9.12^{+0.30}_{-0.17}$  &  $~1.26^{+0.18}_{-0.25}$  &  $ 8.77\pm0.05$  &  $-0.49\pm0.05$  &  $-1.63\pm0.12$  &      \nodata     &  $-2.77\pm0.14$  \\
  J0823+2806  &  0.0472  &  $~9.38^{+0.33}_{-0.19}$  &  $~1.48^{+0.15}_{-0.32}$  &  $ 8.25\pm0.01$  &  $-0.53\pm0.02$  &  $-1.70\pm0.02$  &  $-3.61\pm0.05$  &  $-2.37\pm0.02$  \\
  J0926+4427  &  0.1807  &  $~8.76^{+0.30}_{-0.26}$  &  $~1.03^{+0.13}_{-0.13}$  &  $ 8.05\pm0.09$  &  $-0.56\pm0.12$  &  $-1.66\pm0.12$  &      \nodata     &  $-2.40\pm0.12$  \\
  J0934+5514  &  0.0025  &  $~6.27^{+0.15}_{-0.20}$  &  $-1.52^{+0.09}_{-0.07}$  &  $ 7.09\pm0.01$  &  $ -0.64\pm0.01$     &      \nodata     &      \nodata     &      \nodata     \\
  J0938+5428  &  0.1021  &  $~9.15^{+0.18}_{-0.29}$  &  $~1.05^{+0.20}_{-0.17}$  &  $ 8.22\pm0.02$  &  $-0.59\pm0.03$  &  $-1.63\pm0.03$  &      \nodata     &  $-2.39\pm0.02$  \\
  J0940+2935  &  0.0017  &  $~6.71^{+0.23}_{-0.40}$  &  $-2.01^{+0.42}_{-0.37}$  &  $ 7.99\pm0.09$  &  $-0.73\pm0.12$  &  $-1.83\pm0.10$  &      \nodata     &  $-2.73\pm0.10$  \\
  J0942+3547  &  0.0149  &  $~7.56^{+0.21}_{-0.29}$  &  $-0.76^{+0.19}_{-0.12}$  &  $ 8.00\pm0.09$  &  $-0.59\pm0.02$  &  $-1.63\pm0.02$  &      \nodata     &  $-2.32\pm0.02$  \\
  J0944+3442  &  0.0048  &  $~6.83^{+0.44}_{-0.25}$  &  $-0.78^{+0.19}_{-0.16}$  &  $ 7.66\pm0.22$  &  $-0.57\pm0.14$  &  $-1.39\pm0.13$  &      \nodata     &  $-2.34\pm0.13$  \\
  J0944-0038  &  0.0200  &  $~8.19^{+0.40}_{-0.23}$  &  $-0.01^{+0.28}_{-0.65}$  &  $ 7.82\pm0.01$  &  $-0.70\pm0.01$  &  $-1.66\pm0.02$  &      \nodata     &  $-2.29\pm0.02$  \\
  J1016+3754  &  0.0039  &  $~6.72^{+0.27}_{-0.22}$  &  $-1.17^{+0.18}_{-0.18}$  &  $ 7.57\pm0.04$  &  $-0.68\pm0.02$  &  $-1.72\pm0.02$  &      \nodata     &  $-2.34\pm0.02$  \\
  J1024+0524  &  0.0332  &  $~7.89^{+0.37}_{-0.24}$  &  $~0.21^{+0.14}_{-0.12}$  &  $ 7.89\pm0.01$  &  $-0.67\pm0.01$  &  $-1.76\pm0.02$  &      \nodata     &  $-2.38\pm0.02$  \\
  J1025+3622  &  0.1265  &  $~8.87^{+0.25}_{-0.27}$  &  $~1.04^{+0.14}_{-0.18}$  &  $ 8.19\pm0.02$  &  $-0.62\pm0.03$  &  $-1.79\pm0.04$  &      \nodata     &  $-2.52\pm0.03$  \\
  J1044+0353  &  0.0129  &  $~6.80^{+0.41}_{-0.26}$  &  $-0.59^{+0.11}_{-0.14}$  &  $ 7.55\pm0.02$  &  $-0.70\pm0.01$  &  $-2.12\pm0.01$  &  $-3.93\pm0.01$  &  $-2.54\pm0.01$  \\
  J1105+4444  &  0.0215  &  $~8.98^{+0.29}_{-0.24}$  &  $~0.69^{+0.28}_{-0.22}$  &  $ 8.28\pm0.01$  &  $-0.66\pm0.02$  &  $-1.81\pm0.02$  &  $-3.66\pm0.04$  &  $-2.48\pm0.02$  \\
  J1112+5503  &  0.1316  &  $~9.59^{+0.33}_{-0.19}$  &  $~1.60^{+0.20}_{-0.25}$  &  $ 8.02\pm0.05$  &  $-0.62\pm0.07$  &  $-1.65\pm0.11$  &      \nodata     &  $-2.34\pm0.08$  \\
  J1119+5130  &  0.0045  &  $~6.77^{+0.15}_{-0.28}$  &  $-1.58^{+0.21}_{-0.12}$  &  $ 7.58\pm0.08$  &  $-0.65\pm0.03$  &  $-1.72\pm0.04$  &      \nodata     &  $-2.50\pm0.05$  \\
  J1129+2034  &  0.0047  &  $~8.09^{+0.37}_{-0.27}$  &  $-0.37^{+0.38}_{-0.56}$  &  $ 8.31\pm0.01$  &  $-0.67\pm0.02$  &  $-1.76\pm0.02$  &      \nodata     &  $-2.49\pm0.02$  \\
  J1132+1411  &  0.0176  &  $~7.31^{+0.23}_{-0.26}$  &  $-1.07^{+0.27}_{-0.35}$  &  $ 8.20\pm0.01$  &  $-0.58\pm0.02$  &  $-1.70\pm0.02$  &  $-3.48\pm0.03$  &  $-2.41\pm0.02$  \\
  J1132+5722  &  0.0050  &  $~8.68^{+0.28}_{-0.19}$  &  $~0.44^{+0.24}_{-0.27}$  &  $ 7.34\pm0.02$  &  $-0.62\pm0.04$  &  $-1.48\pm0.06$  &      \nodata     &  $-2.26\pm0.06$  \\
  J1144+4012  &  0.1269  &  $~9.89^{+0.18}_{-0.29}$  &  $~1.51^{+0.20}_{-0.29}$  &  $ 8.65\pm0.08$  &  $-0.45\pm0.15$  &      \nodata     &      \nodata     &  $-2.73\pm0.18$  \\
  J1148+2546  &  0.0451  &  $~8.14^{+0.34}_{-0.24}$  &  $~0.53^{+0.17}_{-0.14}$  &  $ 8.09\pm0.02$  &  $-0.72\pm0.01$  &  $-1.81\pm0.01$  &      \nodata     &  $-2.47\pm0.01$  \\
  J1150+1501  &  0.0024  &  $~6.84^{+0.28}_{-0.30}$  &  $-1.33^{+0.29}_{-0.23}$  &  $ 8.15\pm0.02$  &  $-0.70\pm0.01$  &  $-1.71\pm0.01$  &  $-3.63\pm0.02$  &  $-2.36\pm0.01$  \\
  J1157+3220  &  0.0110  &  $~9.04^{+0.32}_{-0.18}$  &  $~0.97^{+0.21}_{-0.42}$  &  $ 8.44\pm0.03$  &  $-0.69\pm0.04$  &  $-1.76\pm0.04$  &  $-3.55\pm0.06$  &  $-2.50\pm0.04$  \\
  J1200+1343  &  0.0668  &  $~8.12^{+0.47}_{-0.42}$  &  $~0.75^{+0.20}_{-0.16}$  &  $ 8.11\pm0.01$  &  $-0.56\pm0.03$  &  $-1.58\pm0.02$  &      \nodata     &  $-2.28\pm0.02$  \\
  J1225+6109  &  0.0023  &  $~7.12^{+0.34}_{-0.24}$  &  $-1.08^{+0.26}_{-0.26}$  &  $ 8.02\pm0.06$  &  $-0.71\pm0.01$  &  $-1.82\pm0.01$  &  $-3.70\pm0.04$  &  $-2.43\pm0.02$  \\
  J1253-0312  &  0.0227  &  $~7.65^{+0.51}_{-0.23}$  &  $~0.56^{+0.15}_{-0.15}$  &  $ 7.98\pm0.01$  &  $-0.63\pm0.01$  &  $-1.45\pm0.01$  &  $-3.28\pm0.01$  &  $-2.24\pm0.01$  \\
  J1314+3452  &  0.0029  &  $~7.56^{+0.30}_{-0.21}$  &  $-0.67^{+0.23}_{-0.55}$  &  $ 8.27\pm0.23$  &  $-0.67\pm0.01$  &  $-1.79\pm0.01$  &  $-3.85\pm0.03$  &  $-2.48\pm0.01$  \\
  J1323-0132  &  0.0225  &  $~6.31^{+0.26}_{-0.10}$  &  $-0.72^{+0.08}_{-0.09}$  &  $ 7.72\pm0.02$  &  $-0.64\pm0.01$  &  $-1.73\pm0.01$  &      \nodata     &  $-2.23\pm0.01$  \\
  J1359+5726  &  0.0338  &  $~8.41^{+0.31}_{-0.26}$  &  $~0.42^{+0.20}_{-0.14}$  &  $ 8.05\pm0.01$  &  $-0.64\pm0.02$  &  $-1.78\pm0.02$  &      \nodata     &  $-2.44\pm0.02$  \\
  J1416+1223  &  0.1232  &  $~9.59^{+0.32}_{-0.26}$  &  $~1.57^{+0.21}_{-0.25}$  &  $ 8.15\pm0.03$  &  $-0.51\pm0.06$  &  $-1.68\pm0.10$  &      \nodata     &  $-2.47\pm0.08$  \\
  J1418+2102  &  0.0086  &  $~6.22^{+0.49}_{-0.35}$  &  $-1.13^{+0.15}_{-0.16}$  &  $ 7.50\pm0.03$  &  $-0.64\pm0.01$  &  $-1.91\pm0.01$  &  $-3.72\pm0.01$  &  $-2.41\pm0.01$  \\
  J1428+1653  &  0.1817  &  $~9.56^{+0.15}_{-0.23}$  &  $~1.22^{+0.26}_{-0.19}$  &  $ 8.28\pm0.07$  &  $-0.58\pm0.08$  &  $-1.79\pm0.09$  &      \nodata     &  $-2.44\pm0.08$  \\
  J1429+0643  &  0.1735  &  $~8.80^{+0.35}_{-0.21}$  &  $~1.42^{+0.11}_{-0.17}$  &  $ 7.95\pm0.02$  &  $-0.55\pm0.02$  &  $-1.70\pm0.03$  &      \nodata     &  $-2.24\pm0.02$  \\
  J1448-0110  &  0.0274  &  $~7.61^{+0.41}_{-0.24}$  &  $~0.39^{+0.13}_{-0.14}$  &  $ 8.04\pm0.02$  &  $-0.65\pm0.02$  &  $-1.61\pm0.01$  &  $-3.55\pm0.03$  &  $-2.34\pm0.01$  \\
  J1521+0759  &  0.0943  &  $~9.00^{+0.29}_{-0.30}$  &  $~0.95^{+0.16}_{-0.17}$  &  $ 8.66\pm0.05$  &  $-0.37\pm0.14$  &  $-1.62\pm0.31$  &      \nodata     &  $-2.68\pm0.15$  \\
  J1525+0757  &  0.0758  &  $10.10^{+0.28}_{-0.42}$  &  $~1.00^{+0.69}_{-0.24}$  &  $ 8.76\pm0.05$  &  $-0.36\pm0.08$  &  $-1.68\pm0.31$  &      \nodata     &  $-2.66\pm0.14$  \\
  J1545+0858  &  0.0377  &  $~7.52^{+0.43}_{-0.26}$  &  $~0.37^{+0.13}_{-0.17}$  &  $ 7.75\pm0.02$  &  $-0.65\pm0.00$  &  $-1.69\pm0.01$  &      \nodata     &  $-2.31\pm0.01$  \\
  J1612+0817  &  0.1491  &  $~9.78^{+0.28}_{-0.26}$  &  $~1.58^{+0.28}_{-0.24}$  &  $ 8.48\pm0.19$  &  $-0.43\pm0.06$  &  $-1.78\pm0.21$  &      \nodata     &  $-2.49\pm0.10$  \\
\enddata                                
\tablecomments{Columns 2-4: Redshift, stellar mass and SFR taken from \PI, respectively. Column 5: metallicity, and columns 6-9: the Ne/O, S/O, Cl/O and Ar/O abundance ratios. }
\label{tab:abundance_ratios}
\end{deluxetable*} 